\documentclass{mn2e}

\usepackage{times}
\usepackage{graphicx}

\title[The X-ray brightest MACS clusters]{The X-ray brightest clusters of galaxies from the Massive Cluster Survey}

\author[H.\ Ebeling et al.]{\parbox{\textwidth}{H.\ Ebeling$^{1}$, A.C.\ Edge$^{2}$, A.\ Mantz$^{3,4}$, E.\ Barrett$^{1}$, J.\ Patrick Henry$^{1}$,
  C.J.\ Ma$^{1}$, L.\ van Speybroeck$^5$}\\ \\
$^{1}$ Institute for Astronomy, University of Hawaii, 2680
Woodlawn Drive, Honolulu, HI 96822, USA\\
$^{2}$ Department of Physics, University of Durham, South Road, Durham, DH1 3LE, UK\\
$^{3}$ \parbox[t]{0.82\textwidth}{Kavli Institute for Particle Astrophysics and Cosmology at Stanford University, 452 Lomita Mall, Stanford, CA 94305-4085, USA and SLAC National Accelerator Laboratory, 2575 Sand Hill Road, Menlo Park, CA 94025, USA}\\ \vspace*{-2mm} \\
$^4$ NASA Goddard Space Flight Center, Greenbelt, MD 20771, USA\\
$^5$ Harvard-Smithsonian Center for Astrophysics, 60 Garden St, Cambridge, MA 02138, USA}

\begin{document}

\date{submitted December 2009}

\pagerange{\pageref{firstpage}--\pageref{lastpage}} \pubyear{2010}

\maketitle

\label{firstpage}

\begin{abstract}
  We present a statistically complete sample of very X-ray luminous galaxy
  clusters detected in the MAssive Cluster Survey (MACS). This second MACS
  release comprises all 34 MACS clusters with nominal X-ray fluxes in excess of
  $2\times 10^{-12}$ erg s$^{-1}$ cm$^{-2}$ (0.1--2.4 keV) in the ROSAT Bright
  Source Catalogue; two thirds of them are new discoveries. Extending over the
  redshift range from 0.3 to 0.5, this subset complements the complete sample of
  the 12 most distant MACS clusters ($z>0.5$) published in 2007 and further exemplifies
  the efficacy of X-ray selection for the compilation of samples of intrinsically
  massive galaxy clusters. Extensive
  follow-up observations with Chandra/ACIS led to three additional MACS cluster
  candidates being eliminated as (predominantly) X-ray point sources. For
  another four clusters --- which, however, remain in our sample of 34 --- the
  point-source contamination was found to be about 50\%. The median X-ray
  luminosity of $1.3\times 10^{45}$ erg s$^{-1}$ (0.1--2.4 keV, Chandra, within $r_{\rm 500}$) of the
  clusters in this subsample demonstrates the power of the MACS survey strategy
  to find the most extreme and rarest clusters out to significant redshift. A comparison
  of the optical and X-ray data for all clusters in this release finds a wide range of morphologies
  with no obvious bias in favour of either relaxed or merging systems. 
  \end{abstract}


\section{Introduction}

Clusters of galaxies have long been recognised to offer exceptional
opportunities for cosmological and astrophysical studies of remarkable
diversity.  As the largest gravitationally bound entities in the universe, they
are rare objects, originating from extreme overdensities in the primoridal
density field, and growing through continuous accretion as well as serial
mergers into mass concentrations of $10^{14-15}$ $M_{\odot}$ at the present
epoch.

While in-depth studies of individual clusters are invaluable for our
understanding of the physical processes governing the interaction between the
three principal cluster constituents (dark matter, intracluster gas, and
galaxies), it is only through observations of well defined, large samples of
clusters that we can (a) obtain statistically meaningful information about the
properties of clusters as a class of objects evolving on cosmological
timescales, and (b) hope to find exceptional systems, such as, e.g., the Bullet
Cluster (Markevitch et al.\ 2004; Brada\v{c} 2006; Clowe et al,\ 2006), that
enable us to conduct quantitative measurements of fundamental astrophysical
parameters. For many decades, such studies had to be based on optically selected
cluster samples which, while large, have the distinct disadvantage of being
inherently affected and biased by projection effects (van Haarlem, Frenk \&
White 1997; Hicks et al.\ 2008).

%\section{The X-ray advantage}

A nearly unbiased way of selecting statistical cluster samples is through X-ray
surveys, as the X-ray emission, which originates from the diffuse intra-cluster
gas trapped in the clusters' gravitational potential well and heated to virial
temperatures of typically $10^{7-8}$ K, represents direct proof of the existence
of a three-dimensionally bound system. Also, the X-ray emission is much more
peaked at the cluster centre than is the projected galaxy distribution, making
projection effects in X-ray selected cluster samples highly improbable.

\begin{figure*}
\parbox{0.49\textwidth}{
\includegraphics[width=0.49\textwidth]{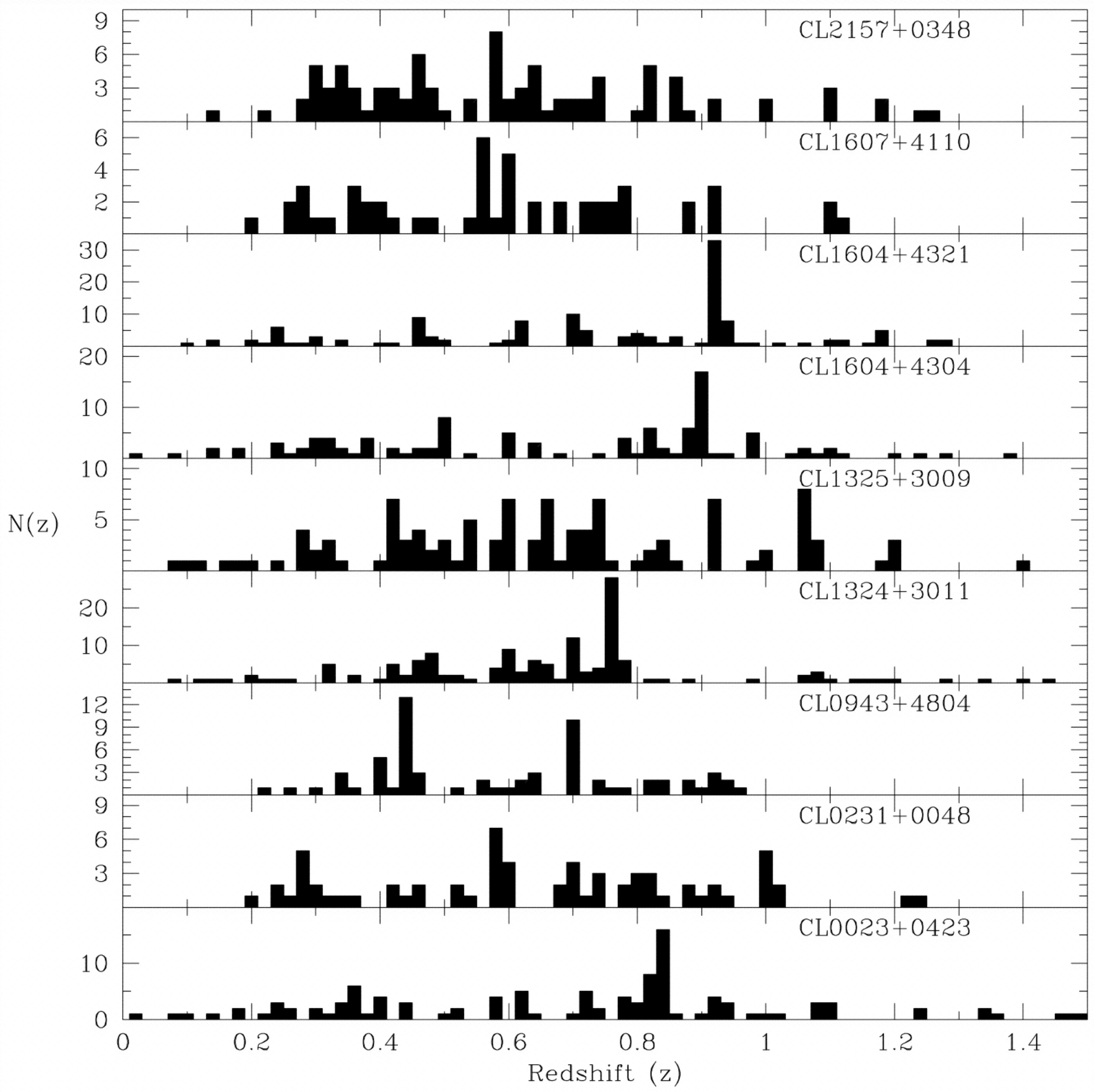}}
\parbox{0.5\textwidth}{
\includegraphics[width=0.51\textwidth]{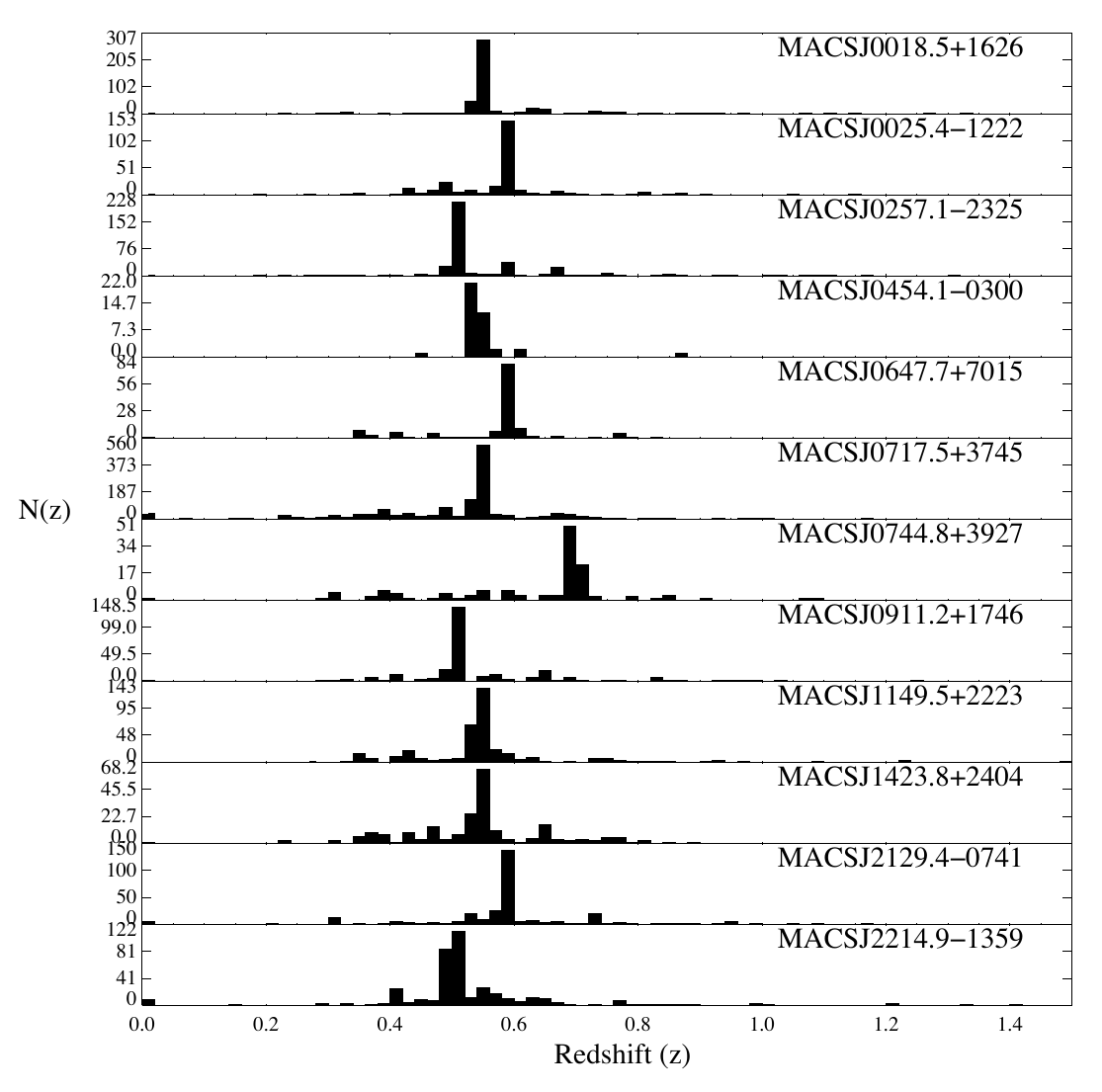}}\hfill
\caption{Histograms of galaxy redshifts in the fields of the nine optically
  selected systems of the PDCS (left; Oke et al.\ 1998) and the twelve most
  distant X-ray selected MACS clusters (right; Ebeling et al.\ 2007). For ease
  of comparison the MACS data are shown over the same redshift range and with
  the same binning as used in the published PDCS figure. Note that both surveys
  used similar criteria to select galaxies for spectroscopic follow-up
  observation.\label{fig:optvsx}}
\end{figure*}

The advantage of X-ray cluster surveys over optical surveys is illustrated in
Fig.~\ref{fig:optvsx} which compares the redshift histograms of the nine
optically selected systems of the Palomar Distant Cluster Survey (PDCS; Oke,
Postman \& Lubin, 1998) with those of the twelve most distant X-ray selected
MACS clusters (Ebeling et al.\ 2007). The severe contamination by fore- and
background structures seen in projection in the PDCS is endemic in optically
selected cluster samples. Pure projection effects like, e.g.,
CL0231+0048 (Fig.~\ref{fig:optvsx}, left) can be largely eliminated by including
information on galaxy colors or redshifts (photometric or spectroscopic) in the
original cluster detection phase. However, even the latest, state-of-the-art optical
cluster samples remain biased, as they are prone to select intrinsically poor
systems whose apparently compact cluster core, high optical richness, and high
velocity dispersion are inflated by line-of-sight alignment and infall (Hicks et
al.\ 2008; Horesh et al.\ 2009). By contrast, X-ray selected cluster samples are
almost entirely free of projection effects since they, by virtue of the X-ray
selection criteria, comprise exclusively intrinsically massive, gravitationally
collapsed systems.

\section{Clusters in the ROSAT All-Sky Survey}

Enormous progress has been made in the past decade in studies of clusters in the
local universe ($z\le 0.3$). The availability of large, representative, X-ray
selected samples compiled from ROSAT All-Sky Survey (RASS, Tr\"umper 1983) data
(Ebeling et al.\ 1996, 1998, 2000; De Grandi et al.\ 1999; Ebe\-ling, Mullis \&
Tully 2002; Cruddace et al.\ 2002; B\"ohringer et al.\ 2004; Kocevski et al.\
2007) has allowed greatly improved, unbiased measurements of the properties of
clusters as an astronomical class of objects.  Especially the ROSAT Brightest
Cluster Sample (BCS, Ebeling et al.\ 1998, 2000) and the REFLEX sample
(B\"ohringer et al.\ 2004) have been used extensively for studies of the local
cluster population (e.g., Allen et al.\ 1992; Crawford et al.\ 1995, 1999;
Ebeling et al.\ 1997; Hudson \& Ebeling 1997; Edge et al.\ 1999, Smith et al.\
2001; Schuecker et al.\ 2001; Allen et al.\ 2003; Kocevski et al.\ 2004, 2006;
Smith et al.\ 2005; Stanek et al.\ 2006; Kocevski \& Ebeling 2006;
Atrio-Barandela et al.\ 2008; Kashlinsky et al.\ 2008).

\begin{figure}
\hspace*{-3mm}\includegraphics[width=0.5\textwidth]{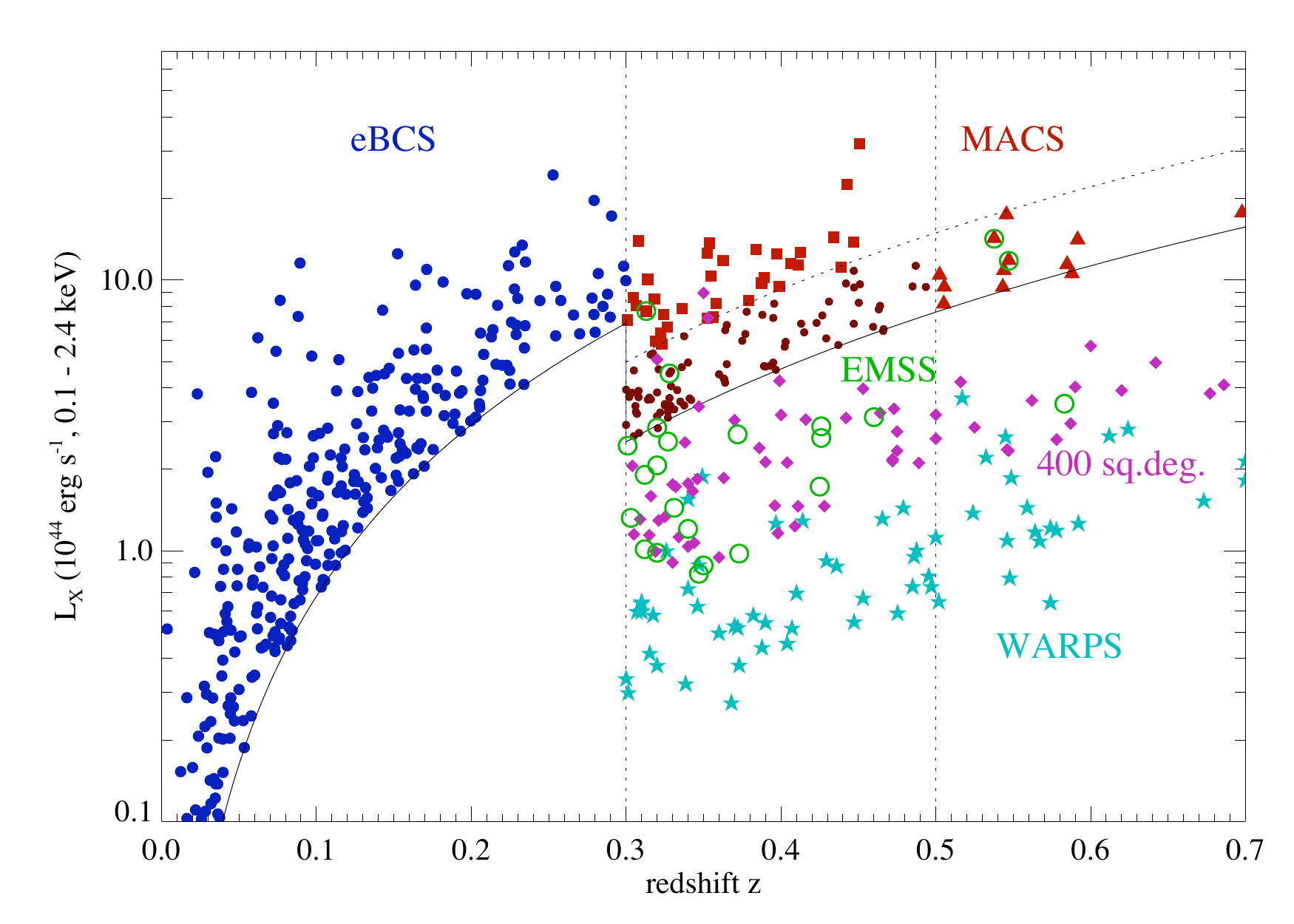}
\caption{$L_{\rm X}$-$z$ distribution of clusters from various X-ray selected samples. By design MACS finds the high-redshift counterparts of the
  most X-ray luminous (and best studied) clusters in the local universe. Note
  also how MACS selects systems that are typically about 10 times more X-ray
  luminous, and thus much more massive, than those found in deeper serendipitous
  cluster surveys such as the EMSS, WARPS, or the 400 square-degree project. Two subsets of the MACS sample are 
  highlighted: the sample presented here (red squares) and the 12 most distant MACS clusters at $z>0.5$ (red triangles; Ebeling et al.\ 2007).  A $\Lambda$CDM cosmology
  ($\Omega_{\rm M}=0.3$, $\Lambda=0.7$, $h_0=0.7$) has been
  assumed. \label{fig:lx-z}}
\end{figure}

At higher redshift, the Massive Cluster Survey (MACS), launched in 1999, has
compiled the first large X-ray selected sample of clusters that are both massive
and distant. Based on sources listed in the RASS Bright Source Catalogue (BSC,
Voges et al.\ 1999) MACS covers the entire extragalactic sky observable from
Mauna Kea ($|b|>20^\circ$, $-40^\circ \leq \delta\leq 80^\circ$), i.e.\ a solid
angle of more than 22,000 deg$^2$, and focuses exclusively on clusters at $z\geq
0.3$.  An overview of the survey strategy is given by Ebeling, Edge \& Henry
(2001); the complete sample of the 12 most distant MACS clusters ($z>0.5$) is
presented in Ebeling et al.\ (2007). Comprising over 120 very X-ray luminous
clusters, the MACS sample represents a 30-fold increase in the number of such
systems known at $z>0.3$. The $L_{\rm X}-z$ distribution of the MACS sample is
shown in Fig.~\ref{fig:lx-z} and compared to that of the BCS, EMSS, WARPS, and 400-sq.-deg.
cluster samples (Ebeling et al.\ 1998, 2000; Gioia \& Luppino 1994; Perlman et
al.\ 2002; Burenin et al.\ 2007; Horner et al.\ 2008; Vikhlinin et al.\ 2009). Multi-wavelength, in-depth follow-up
observations of MACS clusters in particular address a wealth of science issues
across the full spectrum of extragalactic astronomy (LaRoque et al.\ 2003;
Ruderman \& Ebeling 2005; Smail et al.\ 2007; Stott et al.\ 2007, 2009; Ebeling,
Barrett \& Donovan 2004; Kartaltepe et al.\ 2008; Ma et al.\ 2008, 2009;
Brada\v{c} et al.\ 2008; van Weeren et al.\ 2009; Ebeling et al.\ 2009; Bonafede
et al.\ 2009; Smith et al.\ 2009; Limousin et al.\ 2009).

We here present the second complete subsample of MACS clusters comprising the 34
X-ray brightest systems. A $\Lambda$CDM cosmology with $\Omega_m=0.3$,
$\Omega_\Lambda=0.7$ and $H_o=70$\,km\,s$^{-1}$\,Mpc$^{-1}$ is assumed
throughout.

\section{The 34 X-ray brightest MACS clusters}

Two subsets of cluster candidates received special attention in the course of
the compilation of the MACS sample: the most distant ($z>0.5$) and the X-ray
brightest. The former subset is discussed in Ebeling et al.\ (2007); we here
present the second subsample, defined to comprise all MACS clusters with nominal
("detect") fluxes\footnote{Prior to the identification of any RASS BSC sources
  we use an approximate conversion of the net count rate reported in the BSC to
  an unabsorbed X-ray flux by assuming that the observed X-ray emission
  originates from a hot, gaseous plasma with $Z=0.3$ and k$T{=}8$ keV at a
  redshift of $z{=}0.2$.} in the RASS BSC in excess of $2\times 10^{-12}$ erg
cm$^{-2}$ s$^{-1}$ (0.1--2.4 keV).

\subsection{Cluster identification}

Of the 5722 BSC sources meeting the general MACS selection criteria (see Ebeling
et al.\ 2001), 2450 feature detect fluxes above the aforementioned limit. We
identified all of these sources during a 6-year effort which involved repeated
searches of the literature, visual inspection of optical images from public
databases (Digitized Sky Survey and Sloan Digital Sky Survey), scrutiny of
archival data from pointed X-ray observations conducted with ROSAT and Chandra,
and -- as the most time-consuming task -- dedicated imaging and spectroscopic
observations with the University of Hawai`i 2.2m (UH2.2m) telescope and the
Keck-II 10m-telescope.

In recognition of the fact that source identification efforts of this kind are
bound to be subjective to some extent, we took great care not to give undue
weight to the optical appearance of any BSC source. A balance has to be struck
though. Although we did not use "optical richness" as a selection (or cluster
confirmation) criterion, we required the presence of galaxies within the BSC
error circle of approximately 2 arcmin diameter, eliminated sources coinciding
with a single, bright, late-type galaxy, and, during spectroscopic follow-up,
required at least two concordant galaxy redshifts of $z\ge 0.3$. Applying these
optical cluster confirmation criteria in the most conservative fashion led to a
first, tentative sample of 37 MACS clusters with BSC detect fluxes exceeding the
quoted limit. Multi-colour imaging in the V, R, and I passbands was performed
with the UH2.2m telescope of all of these targets in order to allow an
assessment of projection effects, and to enable the identification of potential
stellar counterparts or contaminants of suspicious colour, such as (red) M stars
or (blue) QSOs.

\subsection{Chandra follow-up observations}

MACS was designed to unveil the most X-ray luminous clusters of galaxies at
intermediate redshift and to do so with the least possible bias. Although X-ray
selected, the tentative sample of the 37 X-ray brightest MACS clusters is based
on RASS detections of between 19 and 132 net photons (median: 54), far too few
to allow a secure measurement of any X-ray characteristics beyond estimates of
the source position and flux. Specifically, the existing RASS data do not permit
us to assess whether the respective BSC sources are intrinsically extended and
what fraction of the flux originates from point sources. In order to overcome
both limitations, high-quality X-ray data were obtained with Chandra for all
cluster candidates in this MACS subsample that had not been observed by Chandra
at the time.

All Chandra data were analysed as described in detail in Mantz et al.\
(2010). Specifically, the normalisation of the Chandra fluxes was chosen such
that the latter match those derived from pointed observations of the same
targets with the ROSAT Position Sensitive Proportional Counter (PSPC). This approach is
equivalent to using version 3.5.3 of the CXO calibration database\footnote{Updates to the 
  effective areas used in Chandra CALDB version 3.5.5 break this
  agreement with the PSPC calibration, implying $\sim14$\% higher fluxes than earlier versions (Mantz et al.\ 2010).}.

\begin{figure*}
\vspace*{-8mm}
\includegraphics[width=\textwidth]{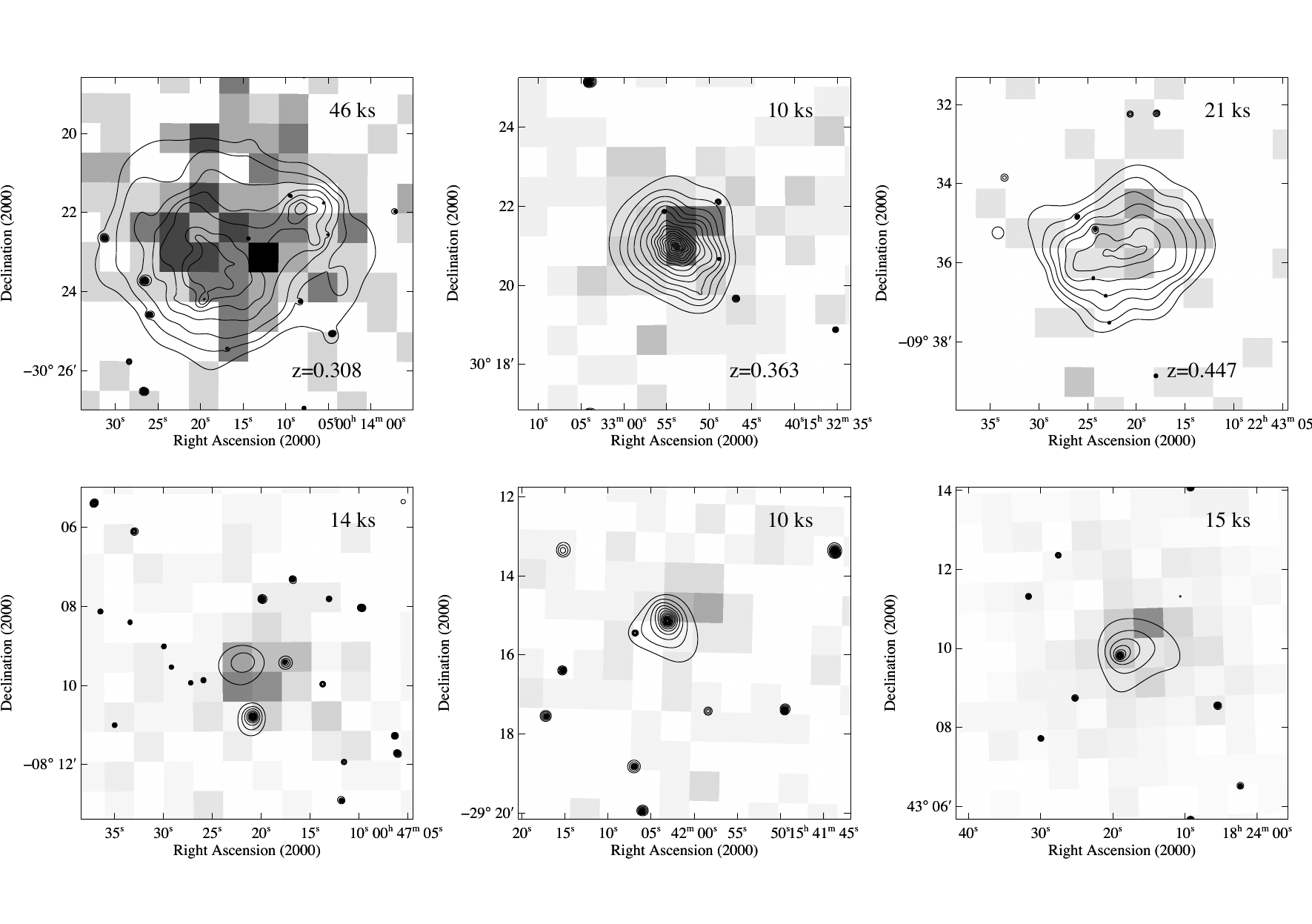}\mbox{}\\*[-8mm]
\caption{Contours of the X-ray surface brightness in the 0.5--7 keV band as
  observed with Chandra/ACIS-I overlaid on the RASS count-rate images (0.1--2.4
  keV) of three confirmed MACS clusters (top row), and of the three erroneous
  identifications found to be dominated by X-ray point sources (bottom row). The
  intensity scaling is linear and the same for all six images; contours are
  spaced logarithmically at the same levels for all images. ACIS-I exposure
  times and cluster redshifts as labeled. The Chandra data were adaptively
  smoothed to $3\sigma$ significance using the {\sc Asmooth} algorithm of
  Ebeling, White, and Rangarajan (2005). Note that, in the RASS, a distant
  virialized cluster like MACSJ1532.8$+$3021 ($z=0.363$, top centre) appears
  just as point-like as the emission from a point source (bottom
  row). \label{fig:rass-cxo}}
\end{figure*}

Figure~\ref{fig:rass-cxo} demonstrates the ability of even very short Chandra
ACIS-I observations (10\,ks) to unambiguously identify the origin and nature of
X-ray emission from cluster candidates at $z\ge 0.3$. We found two of the 37
clusters to contribute less than 10\% of the X-ray flux detected, but not
resolved, in the RASS; a third one (MACSJ1542.0$-$2915) turned out not to be a
cluster at all. The relevant X-ray data for all three of these
misidentifications (subsequently removed from the sample) are shown in the
bottom panel of Fig.~\ref{fig:rass-cxo}. The top panel of the same figure
compares RASS and Chandra data for three confirmed clusters spanning the range
of X-ray fluxes and redshifts of this MACS subsample. Fig.~\ref{fig:rass-cxo}
illustrates two features of the RASS that are critically important for
high-redshift cluster searches: (a) thanks to the low background of the ROSAT
PSPC, even sources consisting of only a few
dozen photons are detected at high significance in the RASS, and (b) dynamically
relaxed clusters at $z>0.3$ appear point-like. Both of these properties lie at
the heart of the MACS strategy to use the faintest RASS sources --- and no
filter for apparent X-ray source extent --- to find the most massive galaxy
clusters out to increasingly high redshift.

\begin{table*}
\centering
\begin{minipage}{\textwidth}
\begin{tiny}
  \caption{Fundamental properties of the 34 X-ray brightest MACS clusters. We
    also list three misidentifications revealed by Chandra observations. All
    cluster redshifts were measured by us, unless noted otherwise. The listed
    coordinates correspond to the peak of the diffuse X-ray emission in our
    Chandra data, except for the X-ray point source MACSJ1542.0$-$2915 for which
    we list the optical position of the QSO. All X-ray fluxes and luminosities are measured
    in the 0.1--2.4 keV band. We list two fluxes determined from
    RASS data: the nominal ``detect flux'' listed in the RASS BSC (to be used
    with the MACS selection function provided in Table~2) and the
    flux within $r_{\rm 500}$ as computed by us from the RASS raw data (see text
    for details). X-ray fluxes, luminosities, and gas temperatures determined
    from Chandra data are also computed within $r_{\rm 500}$ and identical to
    the values listed in Mantz et al.\ (2010). X-ray morphology is assessed visually based on the
    appearance of the X-ray contours and the goodness of the optical/X-ray
    alignment. We use the same simple classification scheme as Ebeling et al.\
    (2007), i.e.\ the assigned morphological classes (from apparently relaxed to
    extremely disturbed) are 1 (pronounced cool core, very good alignment of X-ray
    peak and single cD galaxy), 2 (good optical/X-ray alignment, concentric
    contours), 3 (non-concentric contours, obvious small-scale substructure),
    and 4 (poor optical/X-ray alignment, multiple peaks, no cD galaxy). From the
    differences between classifications made by different authors we estimate
    the uncertainty of the listed values to be less than 1.\label{props}}
\begin{tabular*}{\textwidth}{@{\extracolsep{-1mm}}clccccrrrrrc}
\hline 
MACS name          & other name         & $\alpha$ (J2000) & $\delta$ (J2000) & $z$         &   $n(z)$                  & $f^{\rm X}_{\rm det,BSC}$ & $f^{\rm X}_{\rm r500,RASS}$ & $f^{\rm X}_{\rm r500,CXO}$ & $L^{\rm X}_{\rm r500,CXO}$ & k$T_{\rm CXO}$ (keV) &  morph.\ code\\ \hline \\
MACSJ0011.7$-$1523 &                    & 00 11 42.9 & $-$15 23 22 & 0.379                          & 31             & $2.07 \pm 0.34$ & $ 2.23 \pm 0.41$ & $2.14\pm 0.08$ & $ 8.9\pm 0.3$ & $6.8 \pm 0.6$ & 1\\       
MACSJ0014.3$-$3022 & A2744       & 00 14 18.9 & $-$30 23 22 & \quad0.308\footnote{Struble \& Rood (1999)} & n/a & $5.42 \pm 0.84$ & $ 4.66 \pm 0.51$ & $5.23\pm 0.14$ & $13.6\pm 0.4$ & $8.5 \pm 0.4$ & 4\\
MACSJ0035.4$-$2015 &                    & 00 35 26.5 & $-$20 15 48 & 0.352                          & 34              & $2.05 \pm 0.34$ & $ 2.54 \pm 0.42$ & $3.39\pm 0.18$ & $11.9\pm 0.6$ & $7.3 \pm 0.7$ & 3\\
MACSJ0152.5$-$2852 &                    & 01 52 34.5 & $-$28 53 36 & 0.413                           & 30             & $2.72 \pm 0.33$ & $ 3.20 \pm 0.39$ & $1.69\pm 0.09$ & $ 8.6\pm 0.5$ & $4.7 \pm 0.5$ & 2\\
MACSJ0159.8$-$0849 &                    & 01 59 49.4 & $-$08 49 59 & 0.406                           & 31             & $2.47 \pm 0.34$ & $ 2.53 \pm 0.38$ & $3.37\pm 0.12$ & $16.0\pm 0.6$ & $9.1 \pm 0.7$ & 1\\
MACSJ0242.5$-$2132 &                    & 02 42 35.9 & $-$21 32 26 & \quad0.314\footnote{Wright, Ables \& Allen (1983)}                            & 1            & $3.74 \pm 0.48$ & $ 4.00 \pm 0.57$ & $5.13\pm 0.27$ & $14.2\pm 0.8$ & $5.0 \pm 0.8$ & 1\\
MACSJ0257.6$-$2209 & A402          & 02 57 41.1 & $-$22 09 18 & \quad0.322\footnote{Romer (1994)}      & n/a     & $2.22 \pm 0.41$ & $ 2.83 \pm 0.45$ & $2.43\pm 0.13$ & $ 7.0\pm 0.4$ & $7.0 \pm 0.9$ & 2\\
MACSJ0308.9$+$2645 &                    & 03 08 55.8 & $+$26 45 37 & 0.356                          & 34              & $2.10 \pm 0.40$ & $ 3.36 \pm 0.62$ & $4.16\pm 0.25$ & $14.7\pm 0.9$ & $10.0\pm 1.1$ & 4\\
MACSJ0358.8$-$2955 &                    & 03 58 54.4 & $-$29 55 32 & 0.425                            & 13            & $2.65 \pm 0.31$ & $ 2.81 \pm 0.97$ & $3.60\pm 0.24$ & $18.9\pm 1.2$ & $8.8 \pm 1.1$ & 4\\
\quad MACSJ0404.6$+$1109\footnote{This system is a double cluster: the listed properties refer to the dominant south-western component} &              & 04 04 33.3 & $+$11 07 58 & 0.352 & 1 & $2.27 \pm 0.48$ & $ 2.26 \pm 0.52$ & $1.23\pm 0.14$ & $ 4.3\pm 0.6$ & $7.7 \pm 2.8$ & 4\\
MACSJ0417.5$-$1154 &                    & 04 17 34.7 & $-$11 54 33 & 0.443                             & 41           & $4.13 \pm 0.53$ & $ 4.66 \pm 0.66$ & $5.09\pm 0.27$ & $29.1\pm 1.5$ & $9.5 \pm 1.1$ & 3\\
MACSJ0429.6$-$0253 &                    & 04 29 36.0 & $-$02 53 08 & 0.399                             & 35           & $2.11 \pm 0.42$ & $ 2.68 \pm 0.57$ & $2.35\pm 0.12$ & $10.9\pm 0.6$ & $8.3 \pm 1.6$ & 1\\
MACSJ0520.7$-$1328 &                    & 05 20 42.0 & $-$13 28 50 & 0.336                             & 2           & $2.51 \pm 0.40$ & $ 2.43 \pm 0.47$ & $2.50\pm 0.13$ & $ 7.9\pm 0.4$ & $6.5 \pm 0.8$ & 2\\
MACSJ0547.0$-$3904 &                    & 05 47 01.5 & $-$39 04 26 & 0.319                              & 1          & $2.11 \pm 0.26$ & $ 2.34 \pm 0.30$ & $2.18\pm 0.12$ & $ 6.4\pm 0.4$ & $4.7 \pm 0.5$ & 2\\
MACSJ0947.2$+$7623 & RBS\,0797          & 09 47 13.0 & $+$76 23 14 & 0.354                 & 34                 & $4.16 \pm 0.38$ & $ 4.32 \pm 0.45$ & $5.71\pm 0.30$ & $20.0\pm 1.0$ & $9.5 \pm 2.1$ & 1\\
MACSJ0949.8$+$1708 & Z2661              & 09 49 51.7 & $+$17 07 08 & 0.384                      & 76            & $3.15 \pm 0.43$ & $ 3.66 \pm 0.46$ & $2.55\pm 0.14$ & $10.6\pm 0.6$ & $8.9 \pm 1.8$ & 2\\
MACSJ1115.8$+$0129 &                    & 11 15 52.0 & $+$01 29 55 & 0.355                           & 50             & $2.98 \pm 0.39$ & $ 3.23 \pm 0.48$ & $4.08\pm 0.15$ & $14.5\pm 0.5$ & $9.2 \pm 1.0$ & 1\\
MACSJ1131.8$-$1955 & A1300              & 11 31 54.4 & $-$19 55 42 & 0.306                       & 61                 & $3.15 \pm 0.53$ & $ 4.72 \pm 0.59$ & $5.11\pm 0.28$ & $13.1\pm 0.7$ & $9.4 \pm 1.7$ & 4\\
MACSJ1206.2$-$0847 &                    & 12 06 12.2 & $-$08 48 01 & 0.439                               & 46         & $2.04 \pm 0.39$ & $ 2.87 \pm 0.62$ & $3.79\pm 0.20$ & $21.1\pm 1.1$ & $10.7\pm 1.3$ & 2\\
MACSJ1319.9$+$7003 & A1722              & 13 20 08.4 & $+$70 04 37 & 0.327                     & 53                   & $2.25 \pm 0.26$ & $ 2.66 \pm 0.33$ & $1.41\pm 0.09$ & $ 4.2\pm 0.3$ & $8.4 \pm 2.4$ & 2\\
MACSJ1347.5$-$1144 & RX\,J1347.5$-$1145 & 13 47 30.6 & $-$11 45 10 & 0.451           & 47                             & $5.47 \pm 0.56$ & $ 5.91 \pm 0.69$ & $7.26\pm 0.19$ & $42.2\pm 1.1$ & $10.8\pm 0.8$ & 1\\
MACSJ1427.6$-$2521 &                    & 14 27 39.4 & $-$25 21 02 & 0.318                           & 43             & $3.09 \pm 0.34$ & $ 2.93 \pm 0.77$ & $1.43\pm 0.06$ & $ 4.1\pm 0.2$ & $4.9 \pm 0.6$ & 1\\
MACSJ1532.8$+$3021 & RX\,J1532.9$+$3021 & 15 32 53.8 & $+$30 20 58 & 0.363          & 61                              & $3.58 \pm 0.47$ & $ 4.57 \pm 0.57$ & $5.27\pm 0.19$ & $19.8\pm 0.7$ & $6.8 \pm 1.0$ & 1\\
MACSJ1720.2$+$3536 & Z8201              & 17 20 16.8 & $+$35 36 26 & 0.387                       & 62                 & $2.24 \pm 0.25$ & $ 2.53 \pm 0.32$ & $2.35\pm 0.09$ & $10.2\pm 0.4$ & $7.9 \pm 0.7$ & 1\\
MACSJ1731.6$+$2252 &                    & 17 31 39.1 & $+$22 51 52 & 0.389                             & 82           & $2.36 \pm 0.32$ & $ 2.21 \pm 0.36$ & $2.11\pm 0.12$ & $ 9.3\pm 0.5$ & $5.9 \pm 0.6$ & 4\\
MACSJ1931.8$-$2634 &                    & 19 31 49.6 & $-$26 34 34 & 0.352                              & 35          & $3.65 \pm 0.60$ & $ 4.99 \pm 0.83$ & $5.65\pm 0.30$ & $19.7\pm 1.0$ & $7.5 \pm 1.4$ & 1\\
MACSJ2049.9$-$3217 &                    & 20 49 56.2 & $-$32 16 50 & 0.323                             & 2           & $2.00 \pm 0.40$ & $ 1.96 \pm 0.52$ & $2.10\pm 0.11$ & $ 6.1\pm 0.3$ & $8.1 \pm 1.2$ & 3\\
MACSJ2140.2$-$2339 & MS\,2137.3$-$2353  & 21 40 15.2 & $-$23 39 40 & \quad0.313\footnote{Stocke et al.\ (1991)}  & n/a & $2.86 \pm 0.44$ & $ 3.08 \pm 0.47$ & $4.03\pm 0.14$ & $11.1\pm 0.4$ & $4.7 \pm 0.4$ & 1\\
MACSJ2211.7$-$0349 &                    & 22 11 46.0 & $-$03 49 47 & 0.397                             & 27           & $2.78 \pm 0.47$ & $ 3.25 \pm 0.67$ & $5.39\pm 0.28$ & $24.0\pm 1.2$ & $14.0\pm 2.7$ & 2\\
MACSJ2228.5$+$2036 & RX\,J2228.6$+$2037 & 22 28 34.0 & $+$20 37 18 & 0.411      & 35                                  & $2.26 \pm 0.61$ & $ 2.91 \pm 0.42$ & $2.70\pm 0.15$ & $13.3\pm 0.7$ & $7.4 \pm 0.8$ & 4\\
MACSJ2229.7$-$2755 &                    & 22 29 45.2 & $-$27 55 37 & 0.324                             & 2           & $2.57 \pm 0.42$ & $ 3.72 \pm 0.50$ & $3.37\pm 0.13$ & $10.0\pm 0.4$ & $5.8 \pm 0.7$ & 1\\
MACSJ2243.3$-$0935 &                    & 22 43 21.1 & $-$09 35 43 & 0.447                             & 36           & $2.31 \pm 0.56$ & $ 2.35 \pm 0.67$ & $2.59\pm 0.14$ & $15.2\pm 0.8$ & $8.2 \pm 0.9$ & 3\\
MACSJ2245.0$+$2637 &                    & 22 45 04.6 & $+$26 38 05 & 0.301                            & 1            & $2.88 \pm 0.36$ & $ 2.56 \pm 0.39$ & $3.01\pm 0.16$ & $ 7.6\pm 0.4$ & $5.5 \pm 0.6$ & 1\\
MACSJ2311.5$+$0338 & A2552              & 23 11 33.1 & $+$03 38 07 & 0.305                   & 22                     & $3.48 \pm 0.51$ & $ 3.59 \pm 0.52$ & $5.05\pm 0.40$ & $12.9\pm 1.0$ & $7.5 \pm 1.1$ & 3\\ \hline
MACSJ0047.3$-$0810 &                    & 00 47 21.8 & $-$08 09 25 & 0.317                            & 2            & $3.21 \pm 0.36$ & \multicolumn{5}{l}{AGN (6dF J0047208$-$081046;  $z{=}$0.1532) at 00 47 20.83 $-$08 10 48.5 contributes $>90$\% of BSC flux}\\ 
MACSJ1542.0$-$2915 &                    & \,(15 42 03.10 & \quad$-$29 15 09.7) & -- & n/a & $2.72 \pm 0.43$ & \multicolumn{5}{l}{QSO}\\ 
MACSJ1824.3$+$4309 &                    & 18 24 17.3 & $+$43 09 56 & 0.483 & 12  & $3.19 \pm 0.25$ & \multicolumn{5}{l}{QSO at 18 24 19.01 $+$43 09 49.1 contributes $90$\% of BSC flux}\\ \hline
\end{tabular*}
\end{tiny}
\end{minipage}
\end{table*} 

\subsection{The sample}

\begin{figure*}
\hspace*{-3mm}\includegraphics[width=0.34\textwidth]{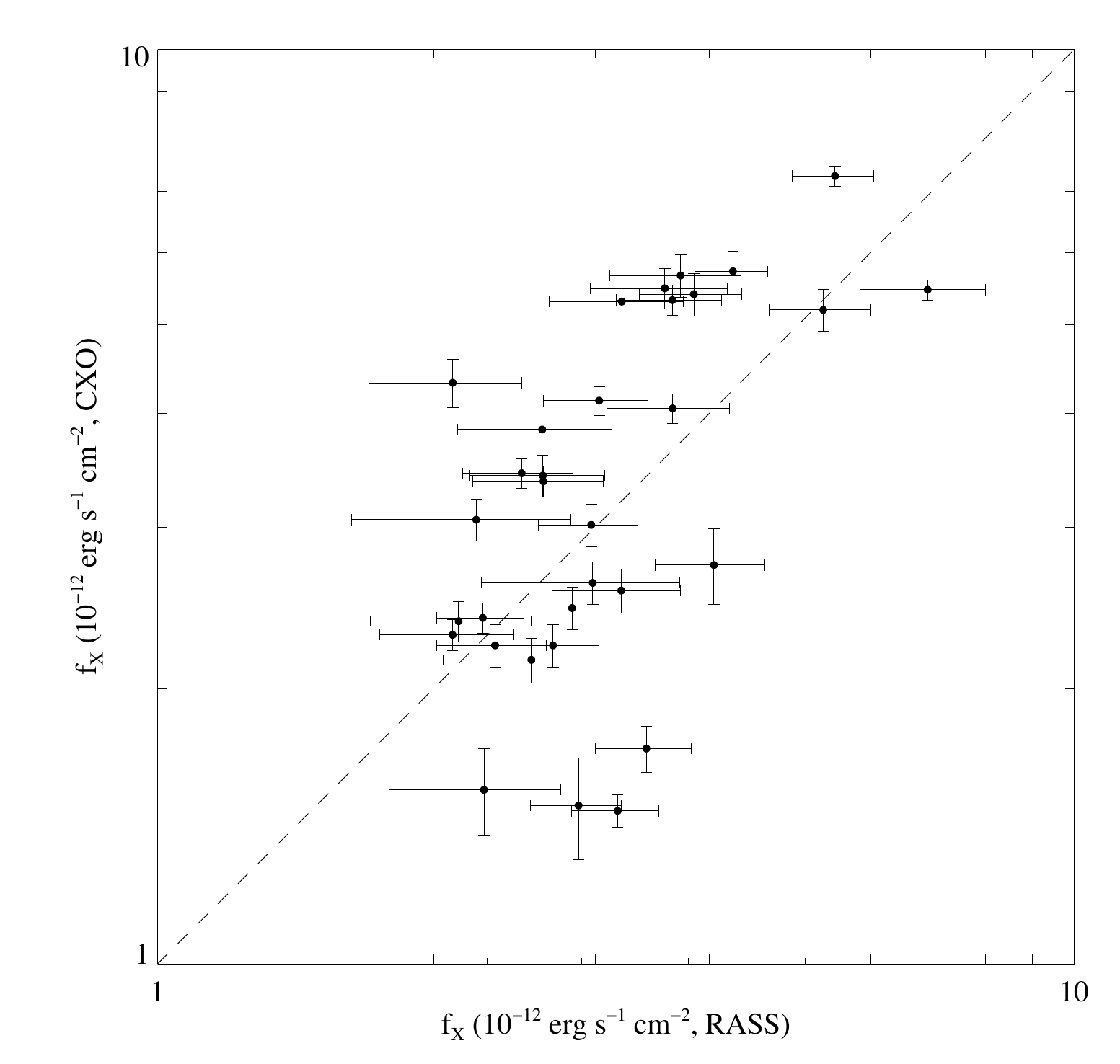}
\hspace*{-3mm}\includegraphics[width=0.34\textwidth]{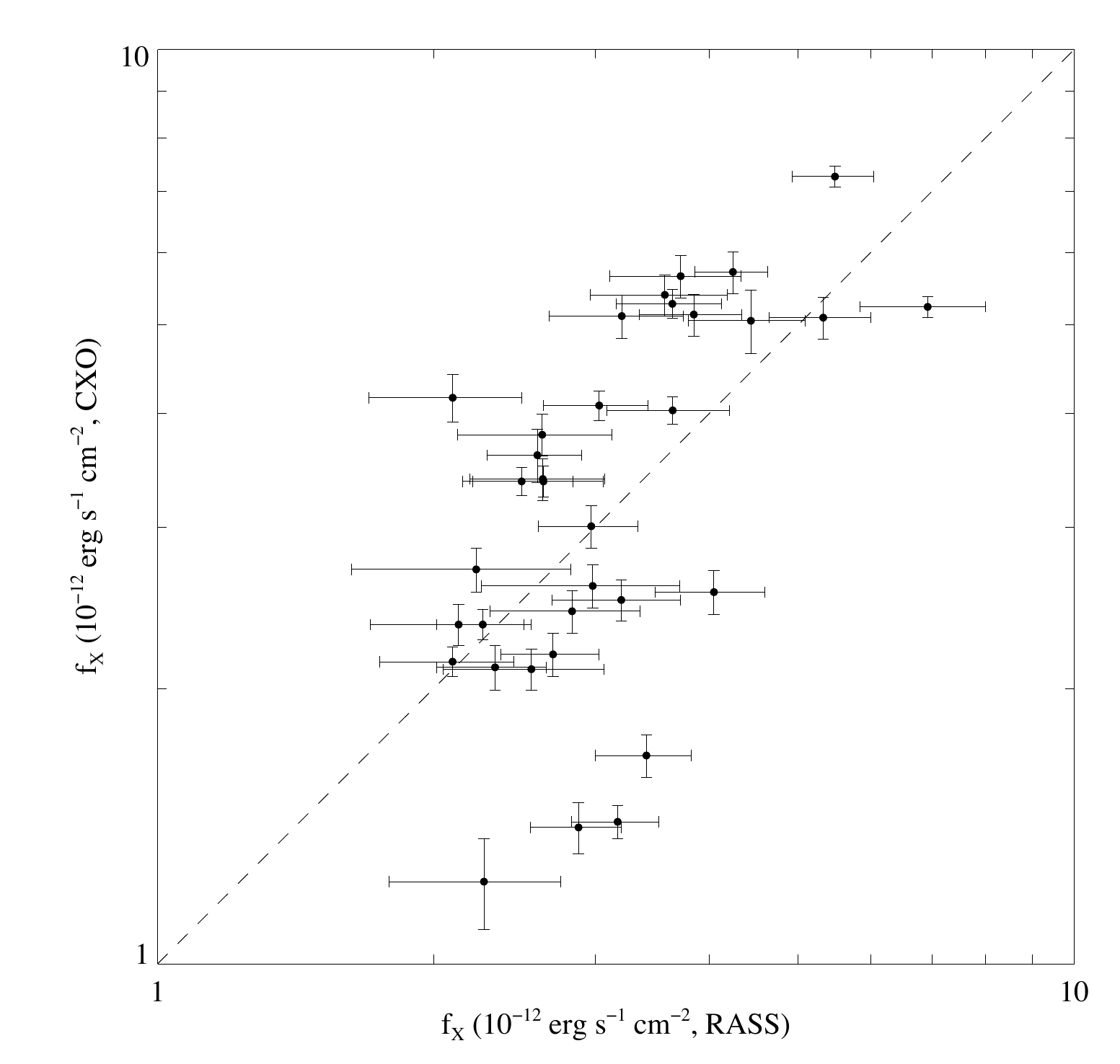}
\hspace*{-3mm}\includegraphics[width=0.34\textwidth]{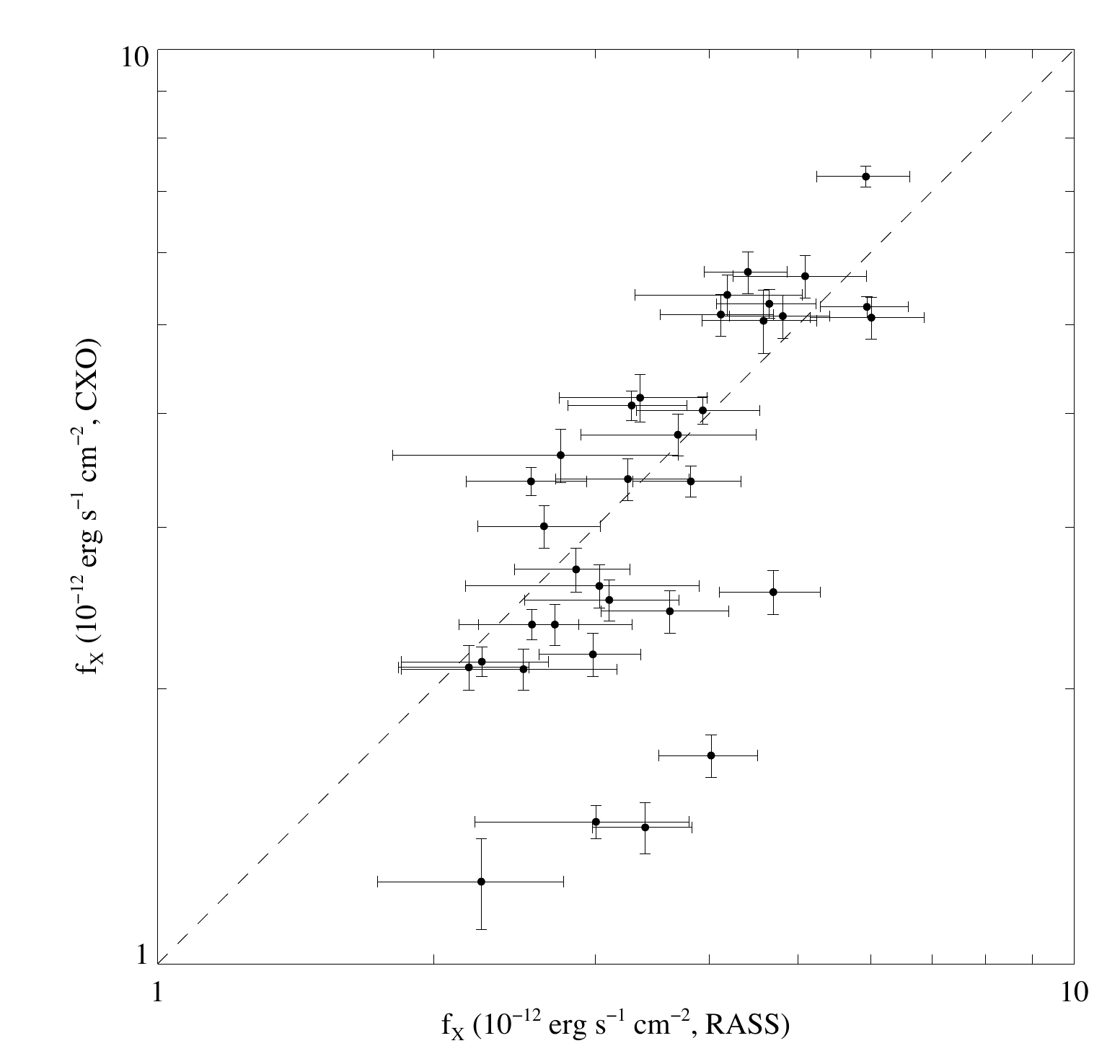}
\caption{X-ray fluxes (0.1--2.4 keV) of the 34 X-ray brightest MACS clusters as
  derived from RASS and Chandra/ACIS observations. The left panel compares the
  Chandra measurements within $r_{\rm 500}$ (including the contribution from X-ray
  point sources) to the RASS values as reported in
  the BSC (but corrected for aperture effects to match
  the Chandra measurement; see text for details). The middle panel shows the same
  comparison except that all X-ray point sources have now been removed from the 
  Chandra measurements. Since point sources contribute, on average, only 3\% of the Chandra flux
  values they are not primarily responsible for the large scatter.
   The right panel, finally, shows a
  visibly improved correlation when the point-source corrected Chandra fluxes are compared to RASS fluxes manually recomputed by us from
  the RASS raw data (within the same BSC detect cells and corrected for the same
  aperture effects). Apart from five outliers discussed in the main body of this paper, the recomputed RASS estimates are, within the errors, consistent with the Chandra
  measurements.  \label{fig:fx-fx}}
\end{figure*}

The availability of Chandra data for all of the X-ray brightest MACS clusters
enables accurate measurements of fundamental cluster properties such as X-ray
flux, X-ray luminosity, and intra-cluster gas temperature. Even more
importantly, and as a first for X-ray cluster surveys, our Chandra data allow us
to remove X-ray point sources prior to the analysis for the entire sample. We list in Table~1 key
properties of the 34 X-ray brightest MACS clusters as well as of the three
cluster candidates found to be dominated by X-ray point sources.

Fig.~\ref{fig:fx-fx} compares the unabsorbed X-ray
fluxes determined from Chandra data to the corresponding estimates from the
RASS. In order to minimize any model-dependent biases we do not extrapolate our
measurements to larger radii to obtain ``total'' fluxes, but plot the Chandra
fluxes measured directly within $r_{\rm 500}$, and the RASS fluxes measured
directly within the BSC detect cell. Note, however, that we slightly adjust the BSC detect fluxes as listed in Table 1 to account for the small difference between $r_{\rm 500}$ (median value for
  this sample: 4.4 arcmin) and the typical radius of the BSC extraction radius
  (5 arcmin), in the process accounting for the RASS point-spread
  function. Although this conversion requires the assumption of a $\beta$ model
  (Cavaliere \& Fusco-Femiano 1976), no significant systematic uncertainties are
  introduced as the resulting corrections are small (median correction:
  4.5\%). A straight comparison of the fluxes (including X-ray point sources) derived from Chandra and
RASS-BSC data (Fig.~\ref{fig:fx-fx}, left) shows more scatter than expected given the size of the error bars. The middle panel of Fig.~\ref{fig:fx-fx} illustrates that the large scatter is not 
caused by X-ray point sources. Although the Chandra data show the diffuse X-ray emission from
five clusters to be only about half as bright as suggested by the BSC fluxes\footnote{For four of these (MACSJ0152.5$-$2852, MACSJ0949.8$+$1708, MACSJ1319.9$+$7003,
and MACSJ1427.6$-$2521), the discrepancy can be attributed to bright X-ray point
sources that fall within the RASS BSC detect cell; the fifth system
(MACSJ0404.6$+$1109) features very extended diffuse emission, only part of which
is captured by the Chandra measurement.}, X-ray point sources are found to contribute, on average, only 3\% to the Chandra flux measurements within $r_{\rm 500}$.
Unable to find other physical causes of the poor agreement, we investigated whether inaccurate
RASS-BSC count rates might be to blame. Indeed, recomputing the source fluxes from
the raw RASS data, within the original BSC detect-cell apertures and using a
local background measured within an annulus extending from 3 to 4 Mpc (radius)
at the cluster redshift, results in a visibly improved correlation with the
Chandra fluxes (Fig.~\ref{fig:fx-fx}, right) and no systematic bias once the
mentioned outliers are excluded ($f_{\rm RASS}/f_{\rm Chandra}=1.0\pm 0.2$). 

\begin{figure}
\hspace*{-3mm}\includegraphics[width=0.5\textwidth]{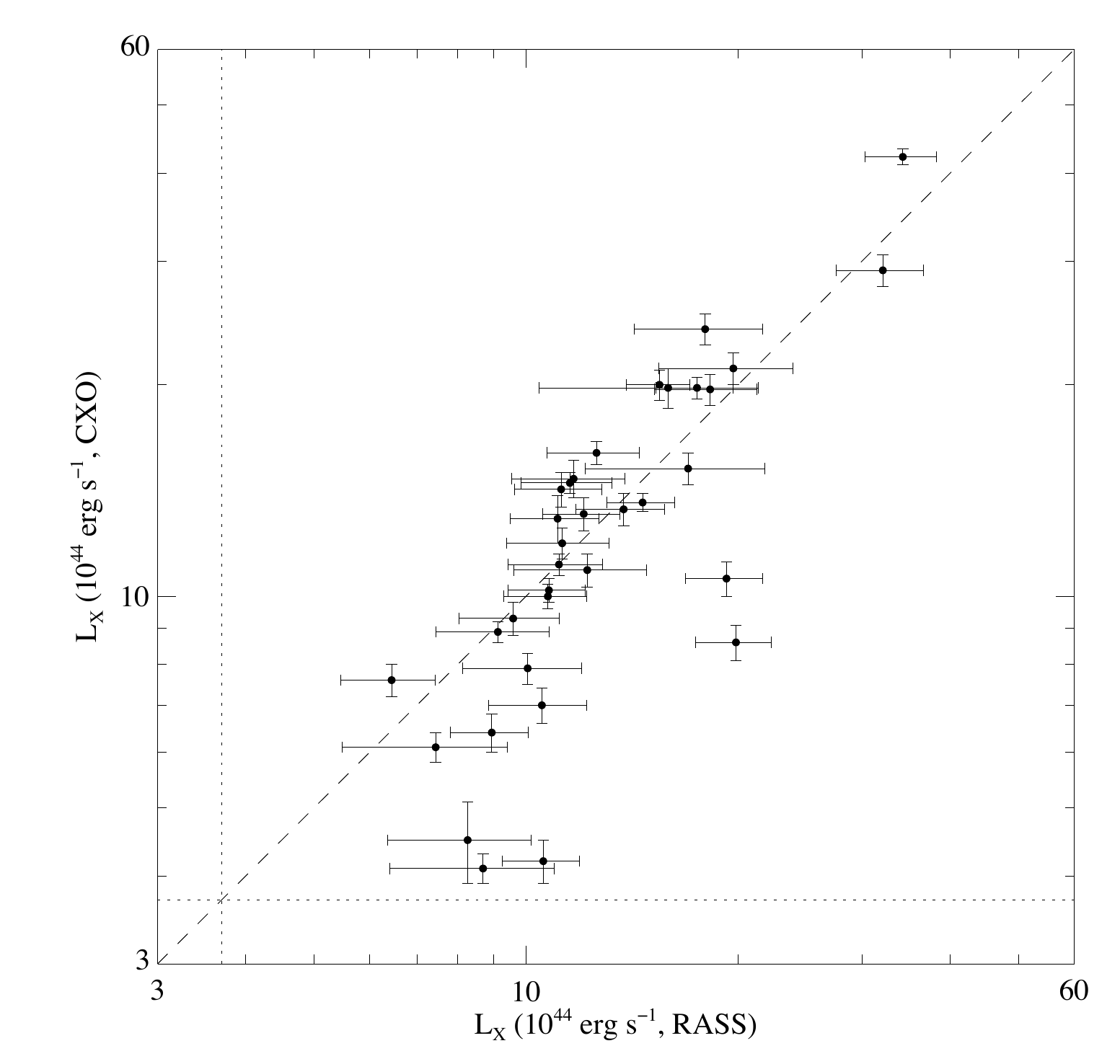}
\caption{Comparison of X-ray luminosities (0.1--2.4 keV, within $r_{\rm 500}$)
  of the 34 X-ray brightest MACS clusters as derived from RASS and Chandra/ACIS
  observations. The RASS values are based on net count rates recomputed by us
  from the RASS raw data (see text for details); the Chandra measurements have
  been corrected for point-source contamination. The dotted lines mark the {\em total}\/ X-ray
  luminosity of the Coma cluster in the same energy band. \label{fig:lx-lx}}
\end{figure}

The ability of the MACS project to find the most X-ray luminous galaxy clusters
out to redshifts of $z\sim 0.5$ and beyond has already been demonstrated by
Ebeling et al.\ (2007) and is confirmed impressively here. We show, in
Fig.~\ref{fig:lx-lx}, a comparison of the estimated RASS X-ray luminosities
(using our recomputed count rates) with the values derived from Chandra
observations.  All systems feature X-ray luminosities (within $r_{\rm 500}$) in
excess of $4\times 10^{44}$ erg s$^{-1}$ (0.1--2.4 keV) after correction for
X-ray point sources, and are thus considerably more X-ray luminous than the Coma
cluster ($L_{\rm X}=3.7 \times 10^{44}$ erg s$^{-1}$, extrapolated to the virial
radius, Ebeling et al.\ 1998). The sample's median X-ray luminosity is
$1.3\times 10^{45}$ erg s$^{-1}$.

Fig.~\ref{fig:opt-cxo} shows overlays of the adaptively smoothed X-ray emission
from all 34 clusters, as observed with Chandra, on colour images created from
optical imaging in the V, R, and I passbands with the UH2.2m telescope in
near-photometric conditions.  This straightforward comparison of cluster
morphologies in the optical and X-ray regime (see final column of Table~1 for a
classification of X-ray morphologies) leads immediately to two conclusions: (a)
MACS is not obviously biased in favour of either 
merging systems or cool-core clusters; (2) a large fraction of the clusters in our
sample, including many systems without obvious cool cores, exhibit excellent
alignment between the location of the brightest cluster galaxy and the peak of
the X-ray emission. We discuss these findings in more detail in a separate paper
(Mann \& Ebeling, in preparation).

\begin{figure*}
\includegraphics[type=pdf,ext=.pdf,read=.pdf,width=0.33\textwidth]{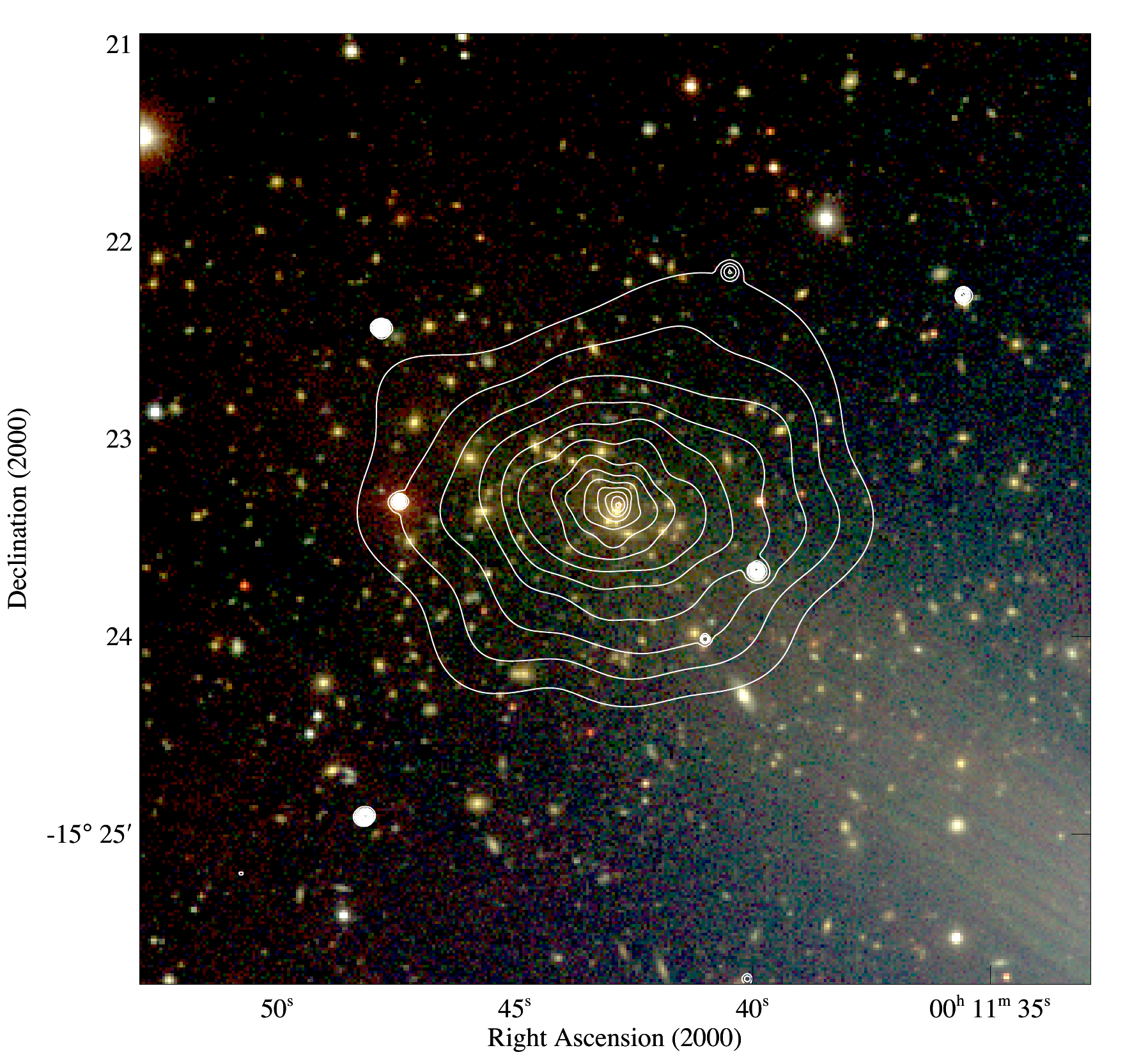}
\includegraphics[type=pdf,ext=.pdf,read=.pdf,width=0.33\textwidth]{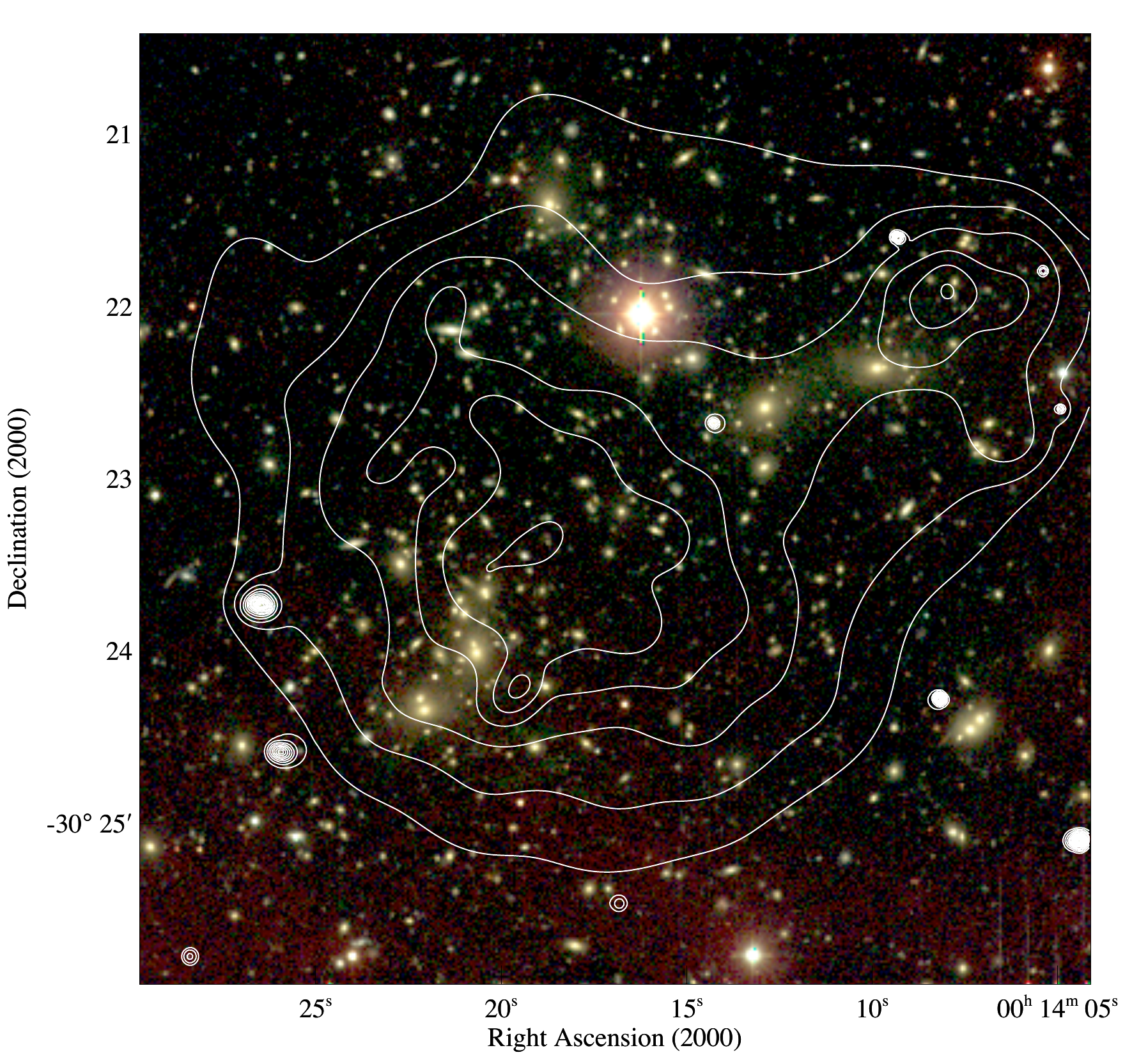}
\includegraphics[type=pdf,ext=.pdf,read=.pdf,width=0.33\textwidth]{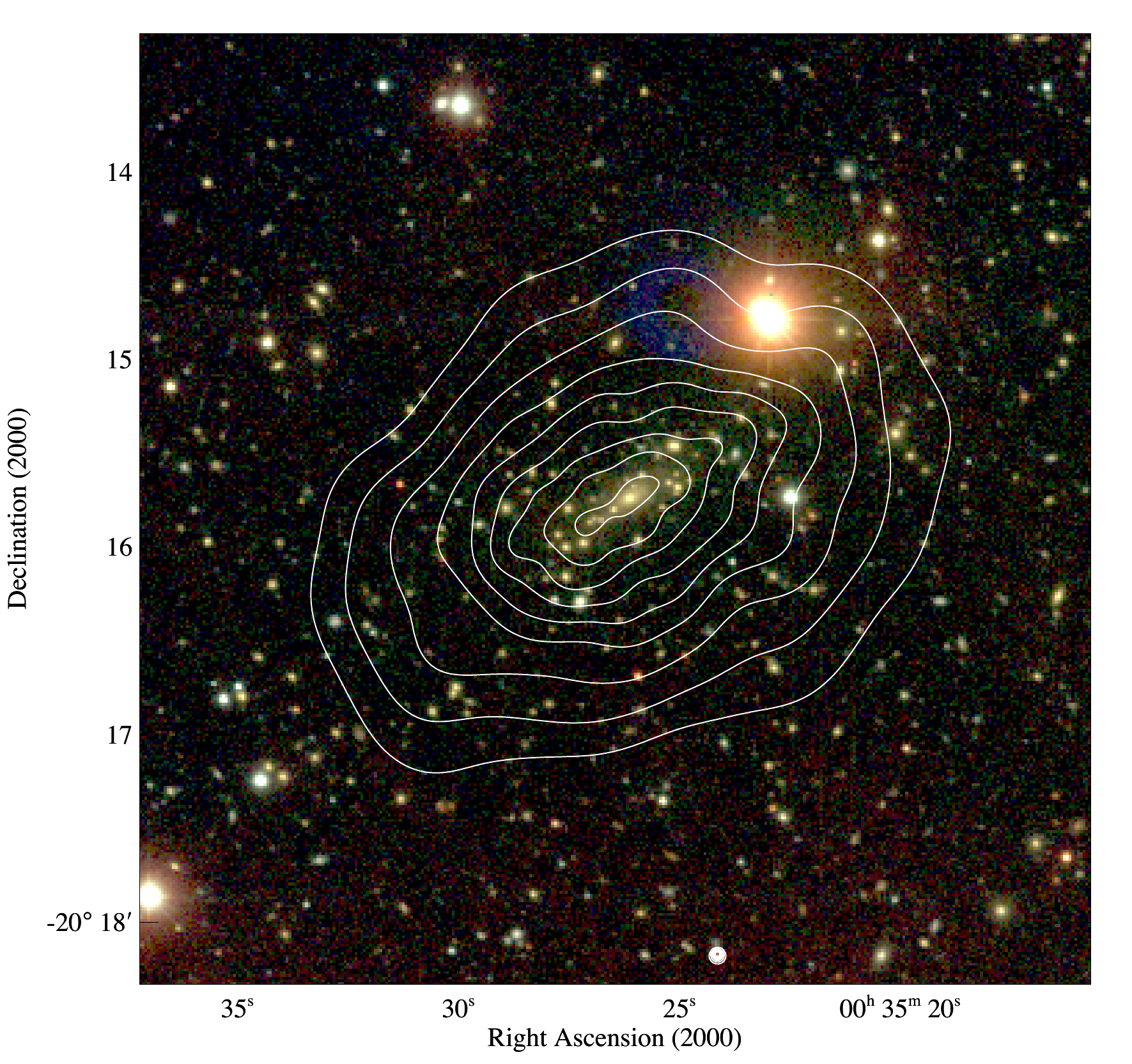}
\includegraphics[type=pdf,ext=.pdf,read=.pdf,width=0.33\textwidth]{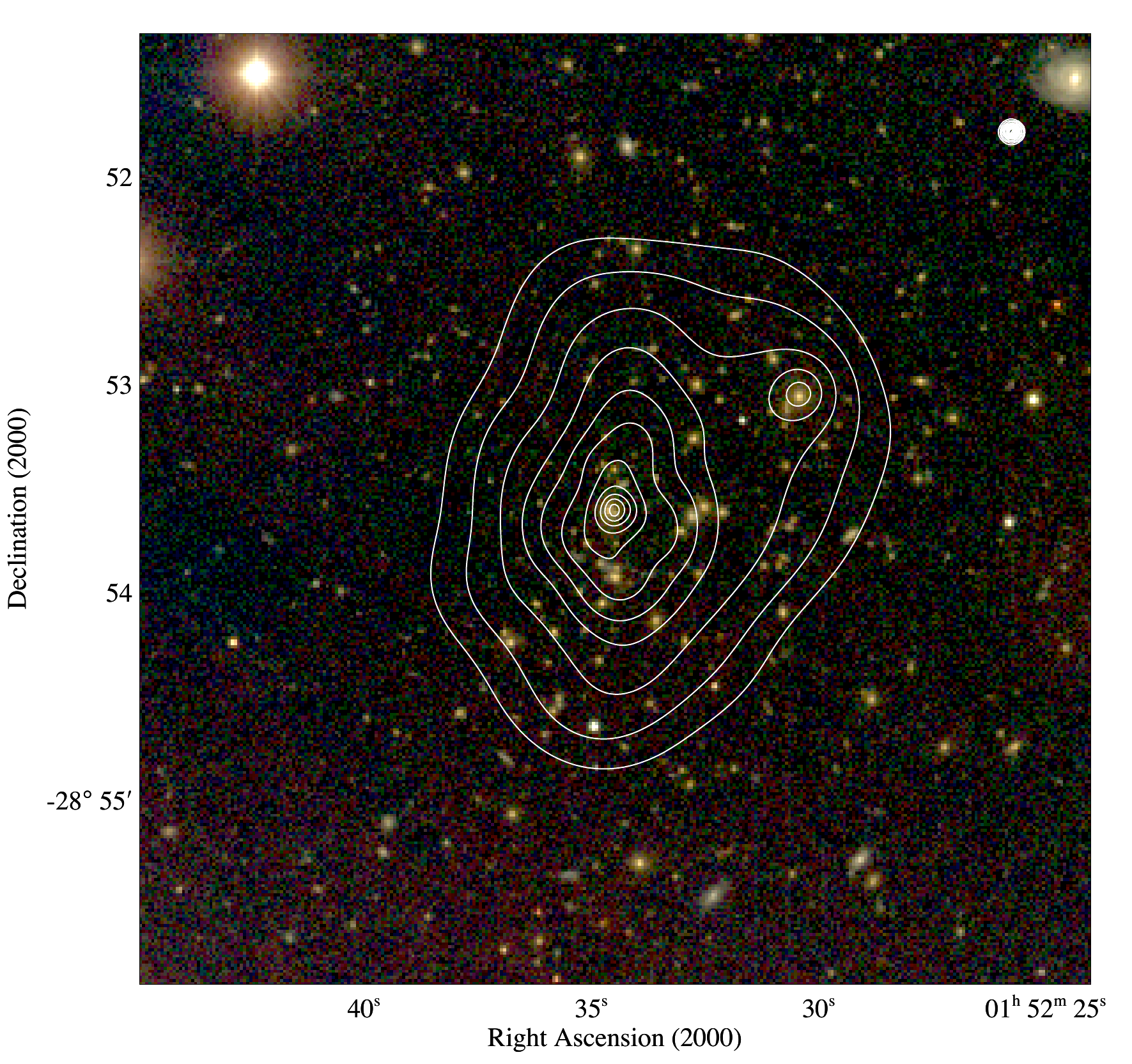}
\includegraphics[type=pdf,ext=.pdf,read=.pdf,width=0.33\textwidth]{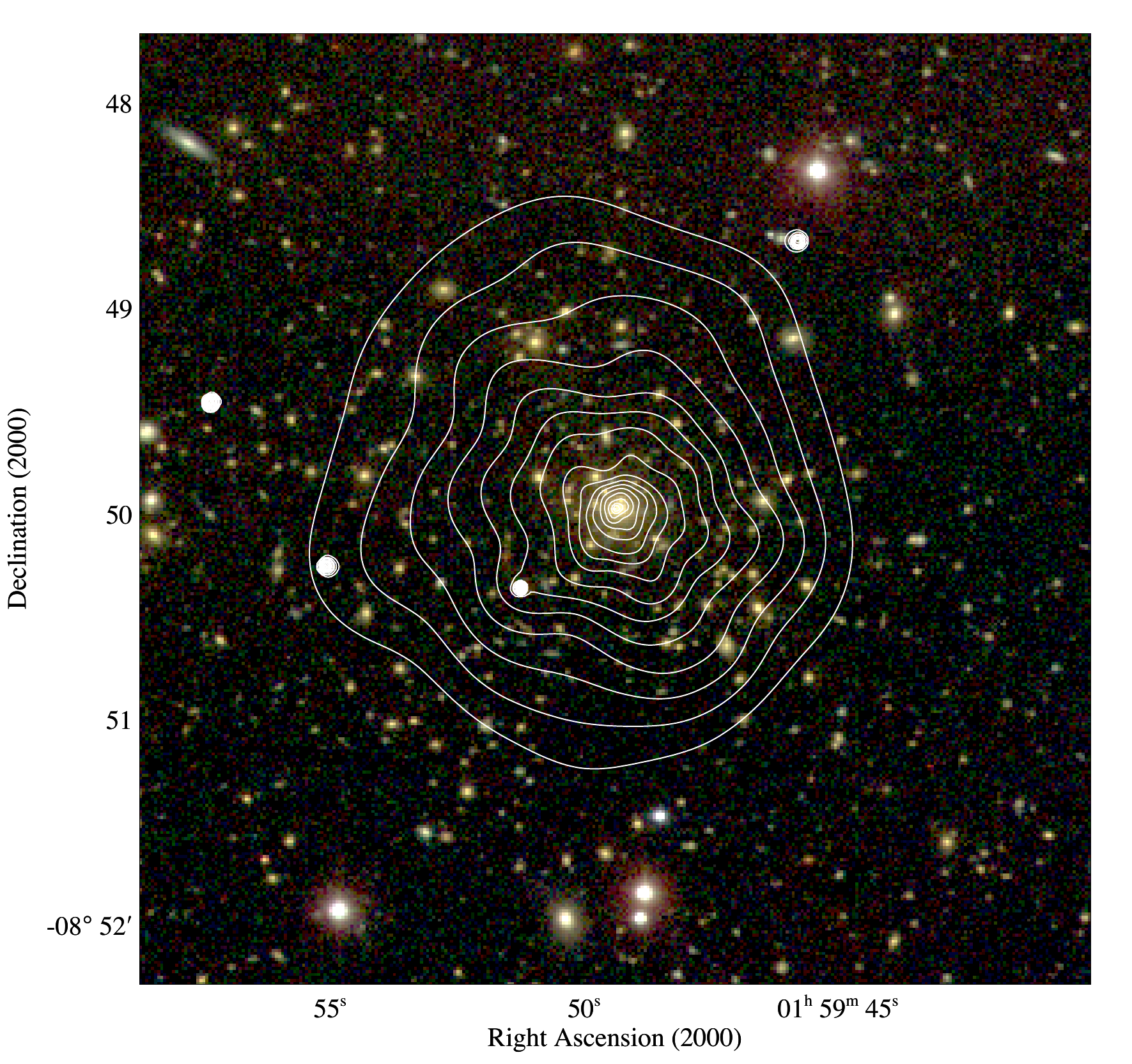}
\includegraphics[type=pdf,ext=.pdf,read=.pdf,width=0.33\textwidth]{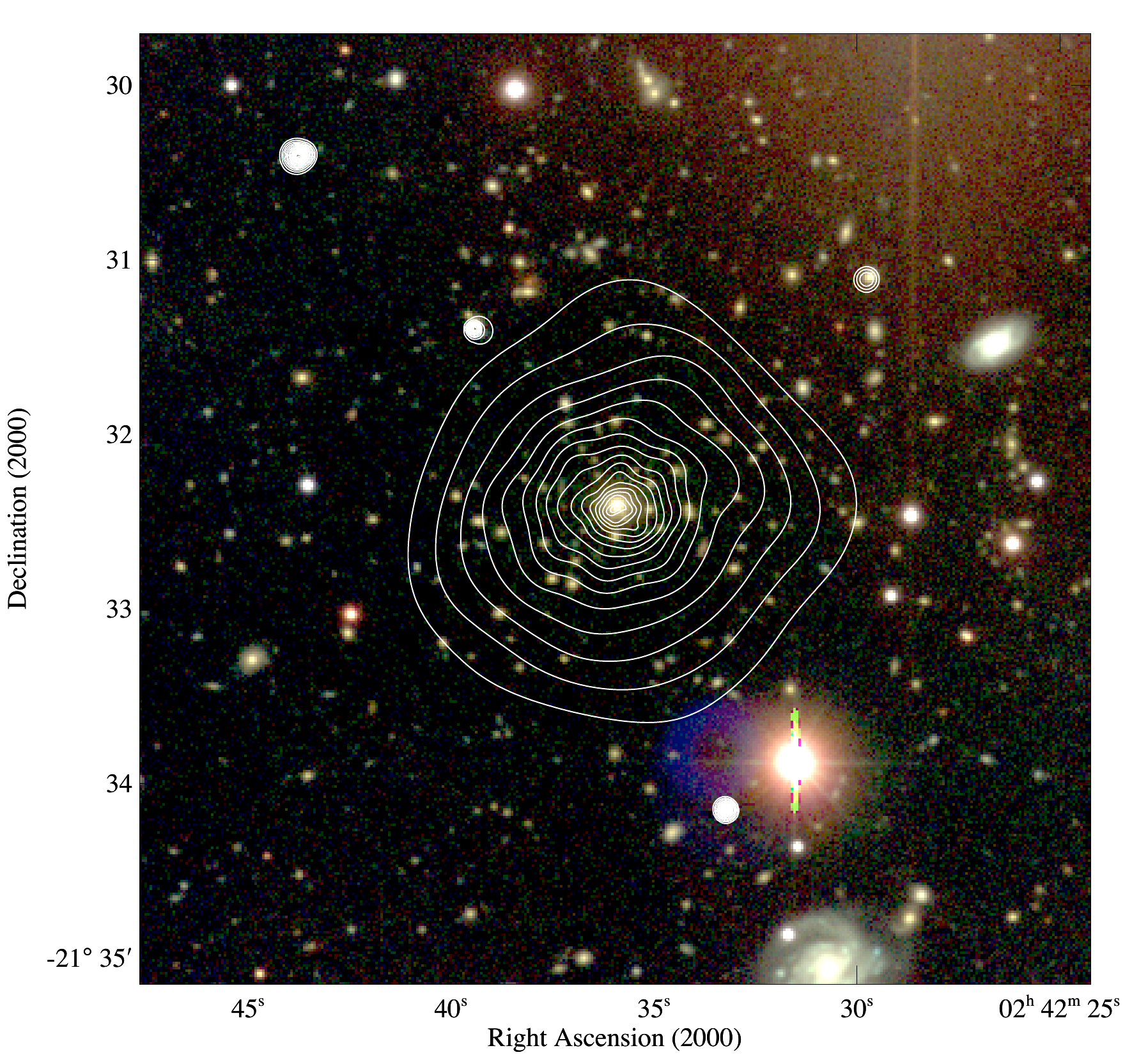}
\includegraphics[type=pdf,ext=.pdf,read=.pdf,width=0.33\textwidth]{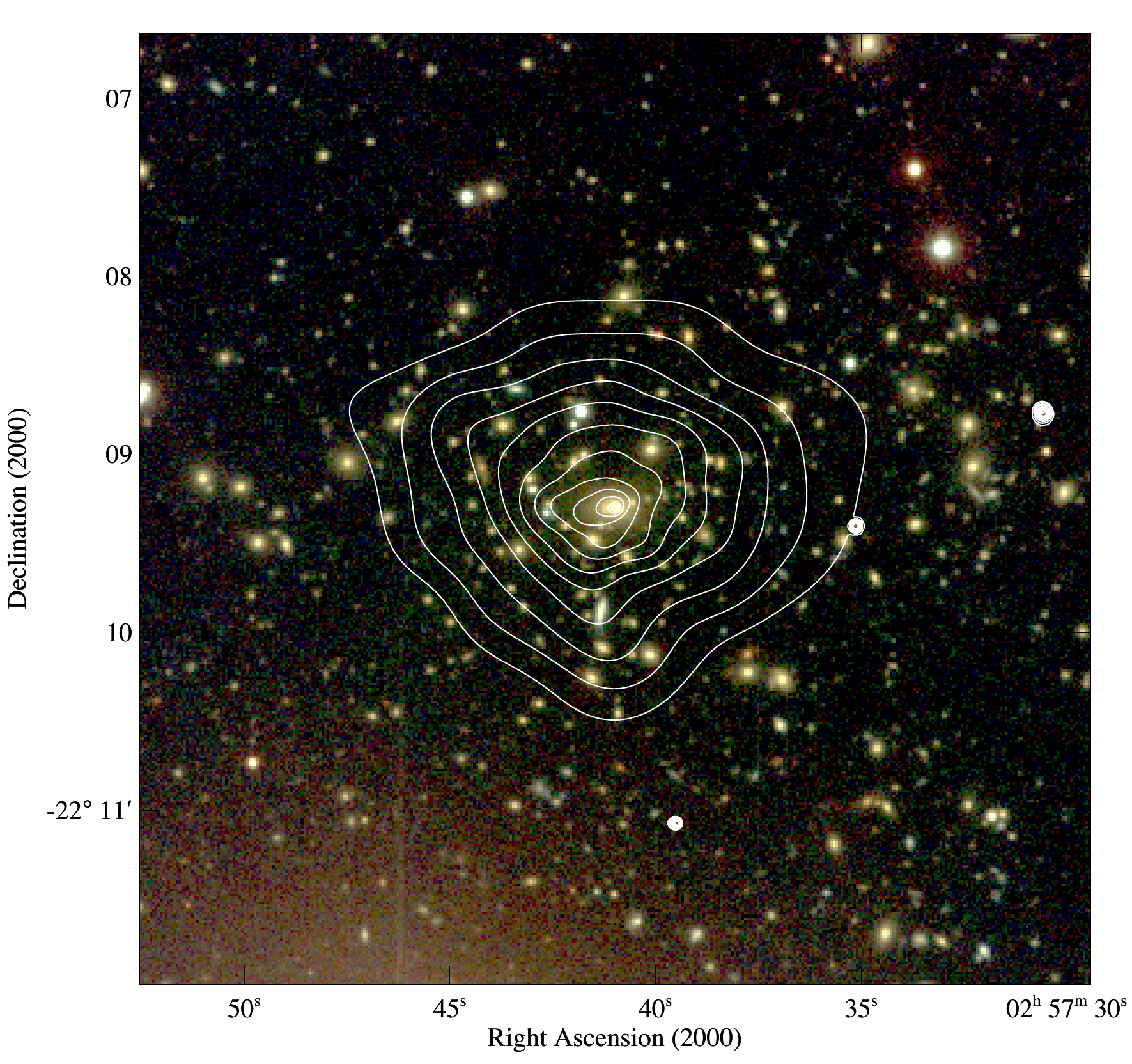}
\includegraphics[type=pdf,ext=.pdf,read=.pdf,width=0.33\textwidth]{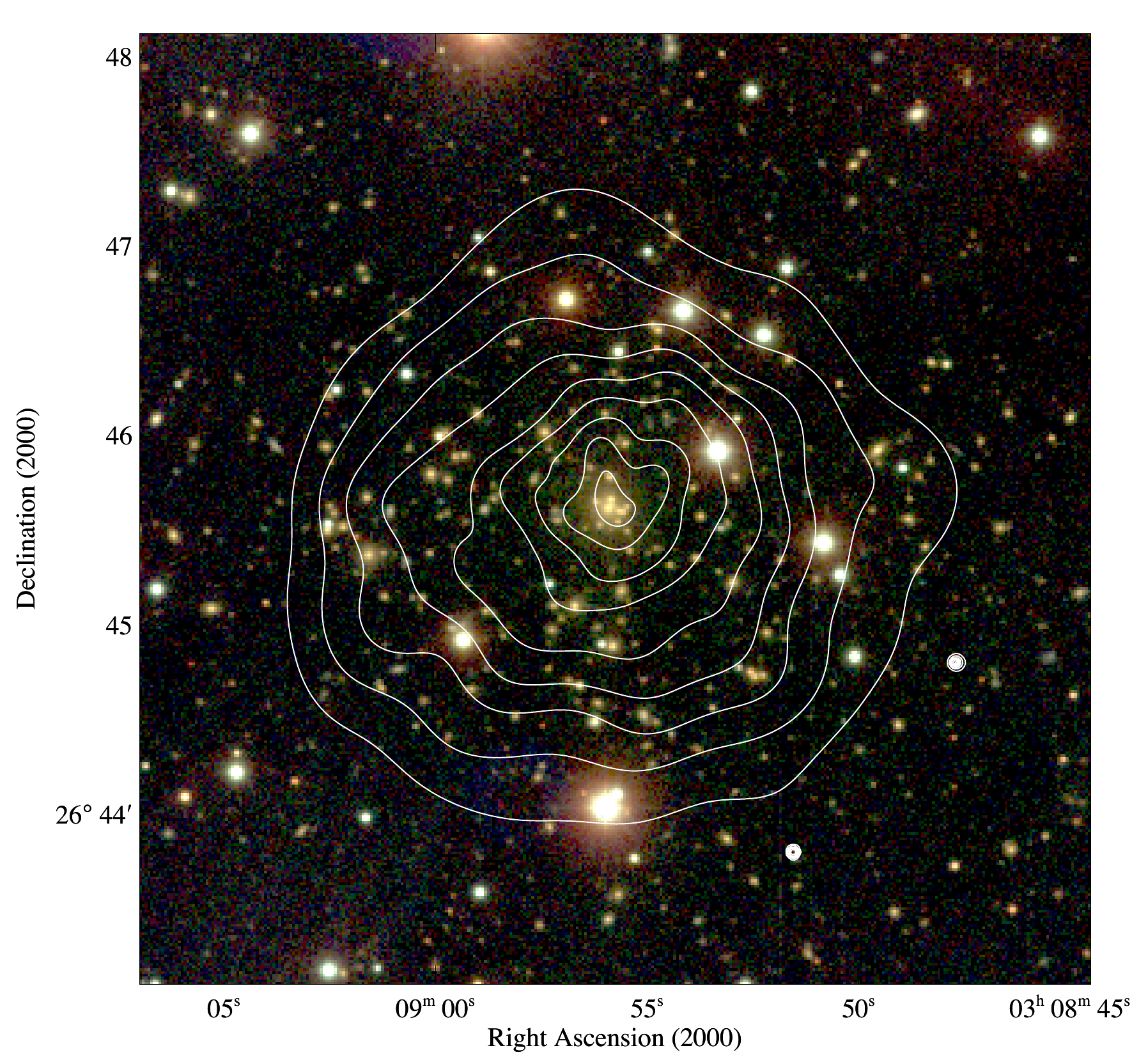}
\includegraphics[type=pdf,ext=.pdf,read=.pdf,width=0.33\textwidth]{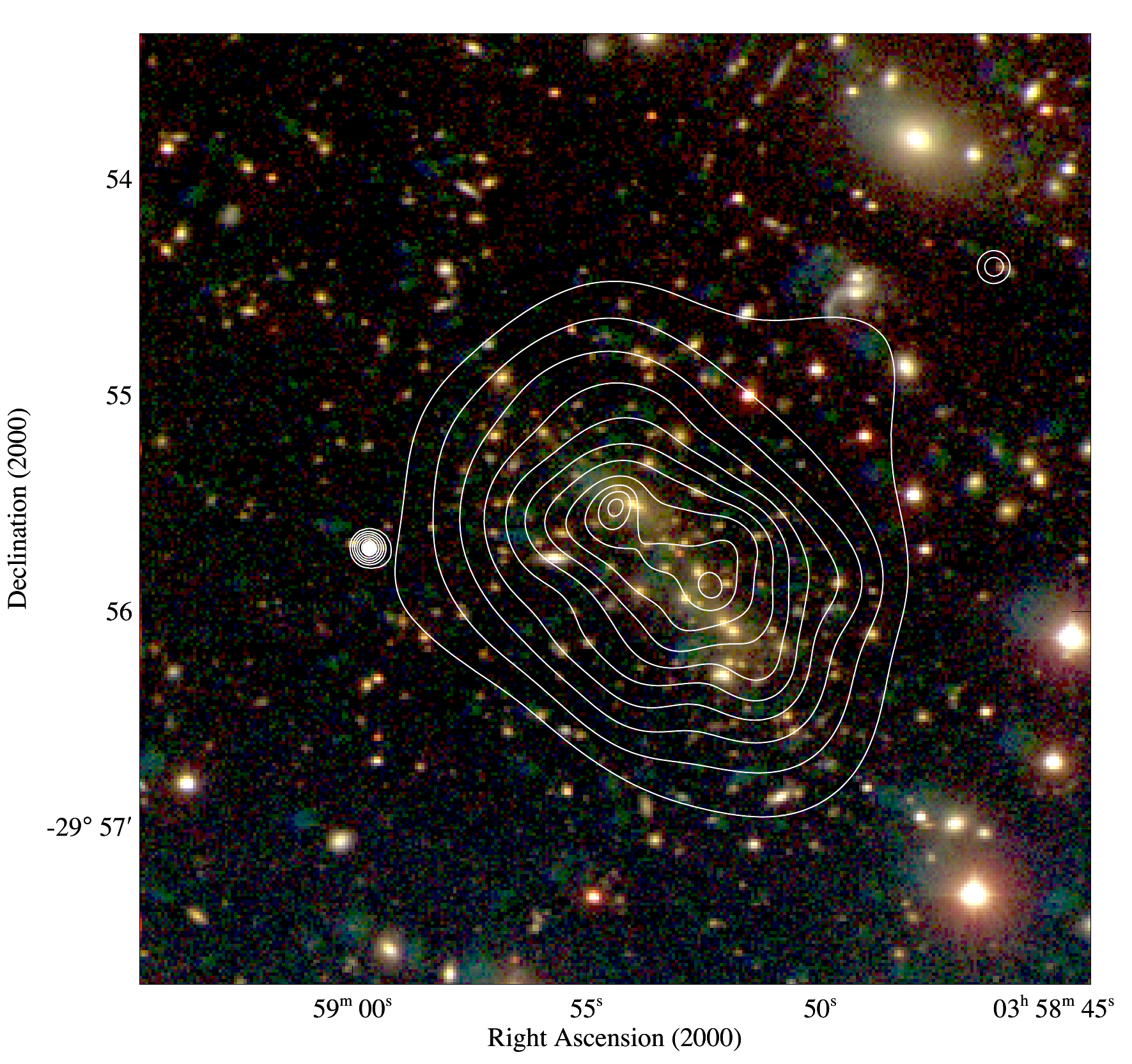}
\includegraphics[type=pdf,ext=.pdf,read=.pdf,width=0.33\textwidth]{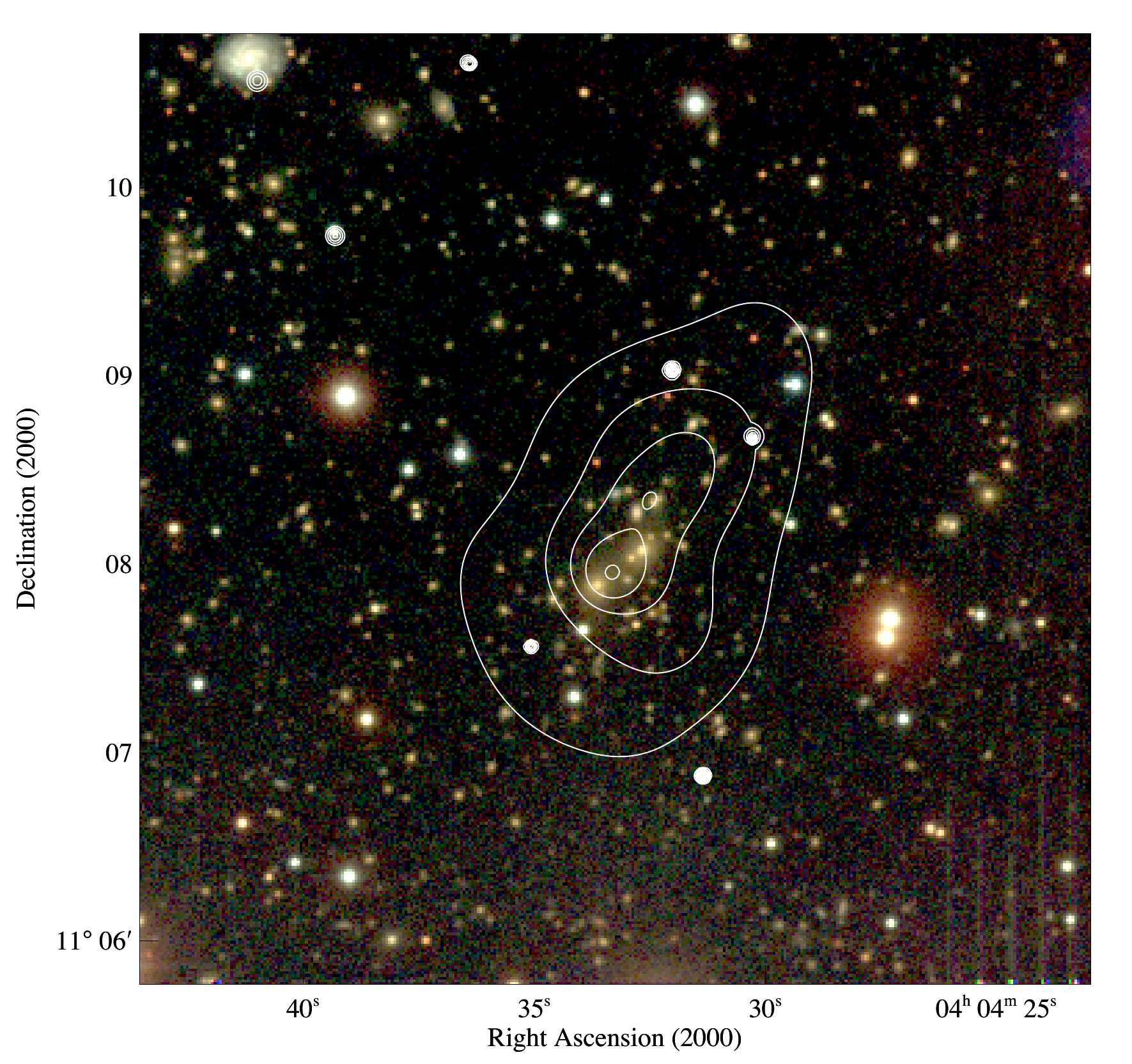}
\includegraphics[type=pdf,ext=.pdf,read=.pdf,width=0.33\textwidth]{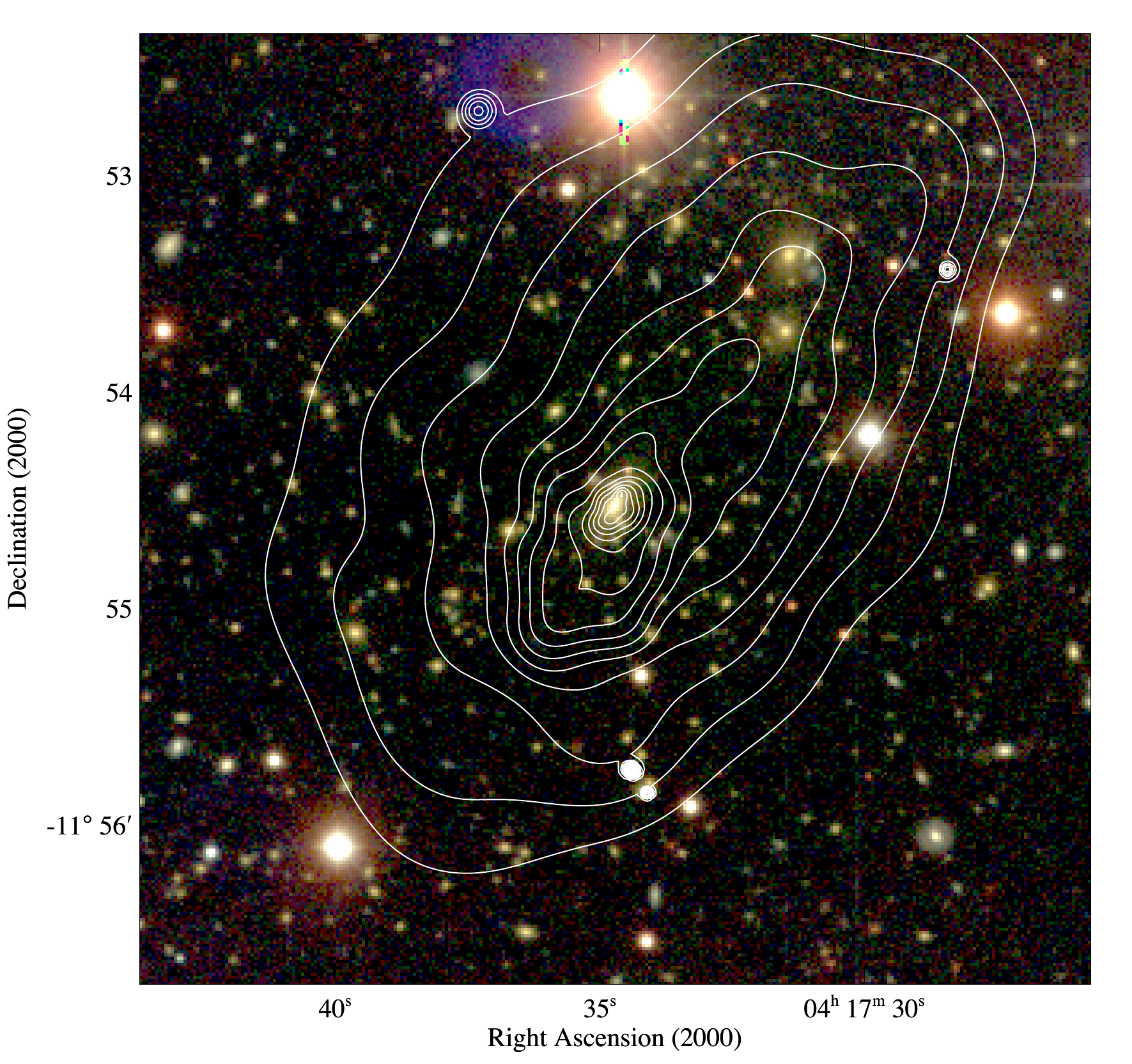}
\includegraphics[type=pdf,ext=.pdf,read=.pdf,width=0.33\textwidth]{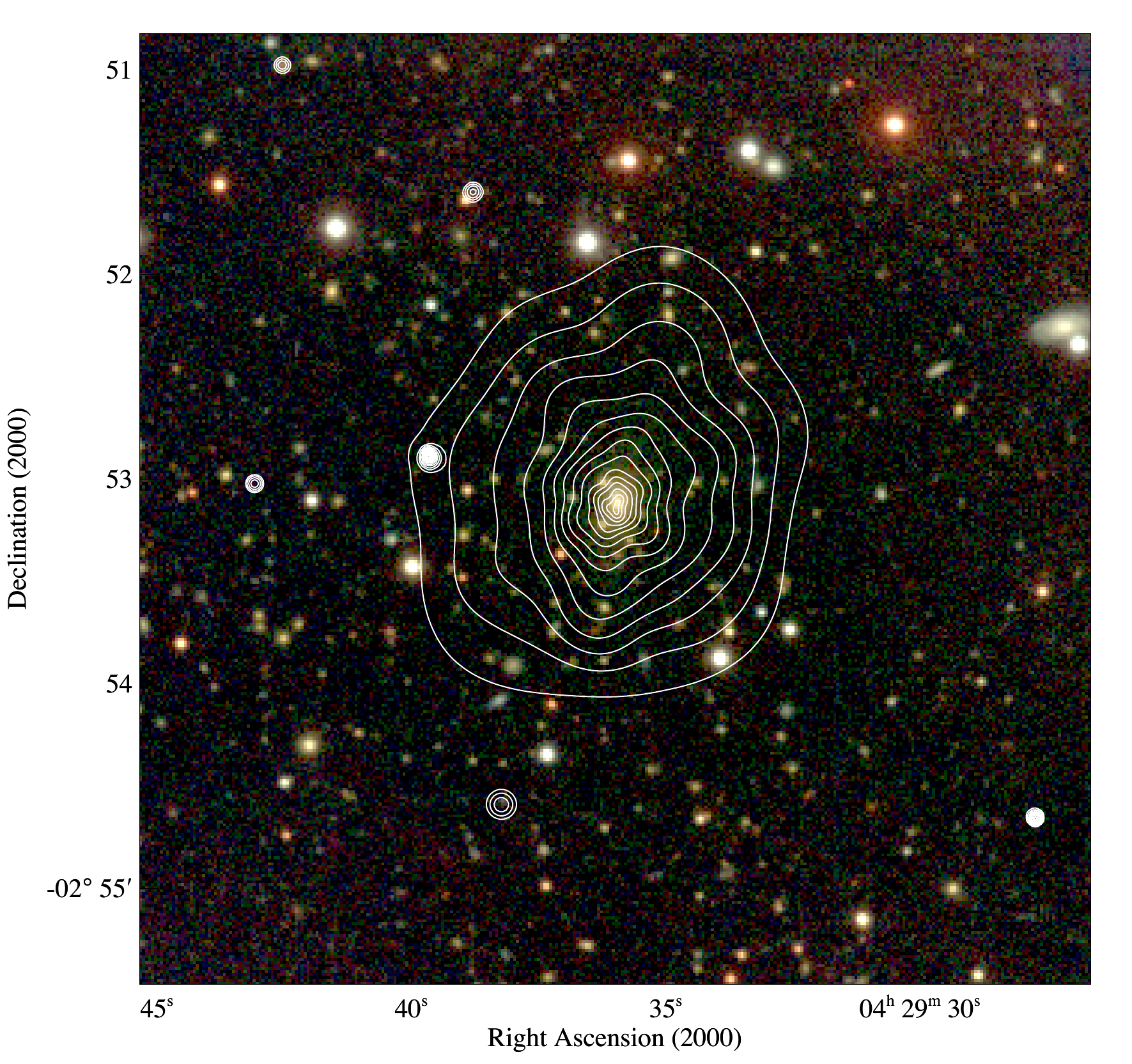}
\caption{Contours of the X-ray surface brightness in the 0.5--7 keV band as
  observed with Chandra/ACIS-I overlaid on colour images obtained with the
  UH2.2m telescope (V,R,I; 12\,min per filter). All images span 1.5 Mpc on the side at the cluster redshift. Contours are spaced
  logarithmically at the same levels for all images, except for MACSJ2140.2--2339 for which only ACIS-S data are available; we omit the lowest two contour levels to account for the higher background of the ACIS-S detector. The Chandra data were
  adaptively smoothed to $3\sigma$ significance using the {\sc Asmooth}
  algorithm of Ebeling, White, and Rangarajan (2005). The final two panels (framed in black and spanning 4.5 arcmin on the side) show
  two of the three candidates revealed as point sources by Chandra and thence
  removed from the sample (for the third candidate listed in Table~1 and shown in Fig.~3 we do not have a UH2.2m colour image). \label{fig:opt-cxo}}
\end{figure*}

\begin{figure*}
\includegraphics[type=pdf,ext=.pdf,read=.pdf,width=0.33\textwidth]{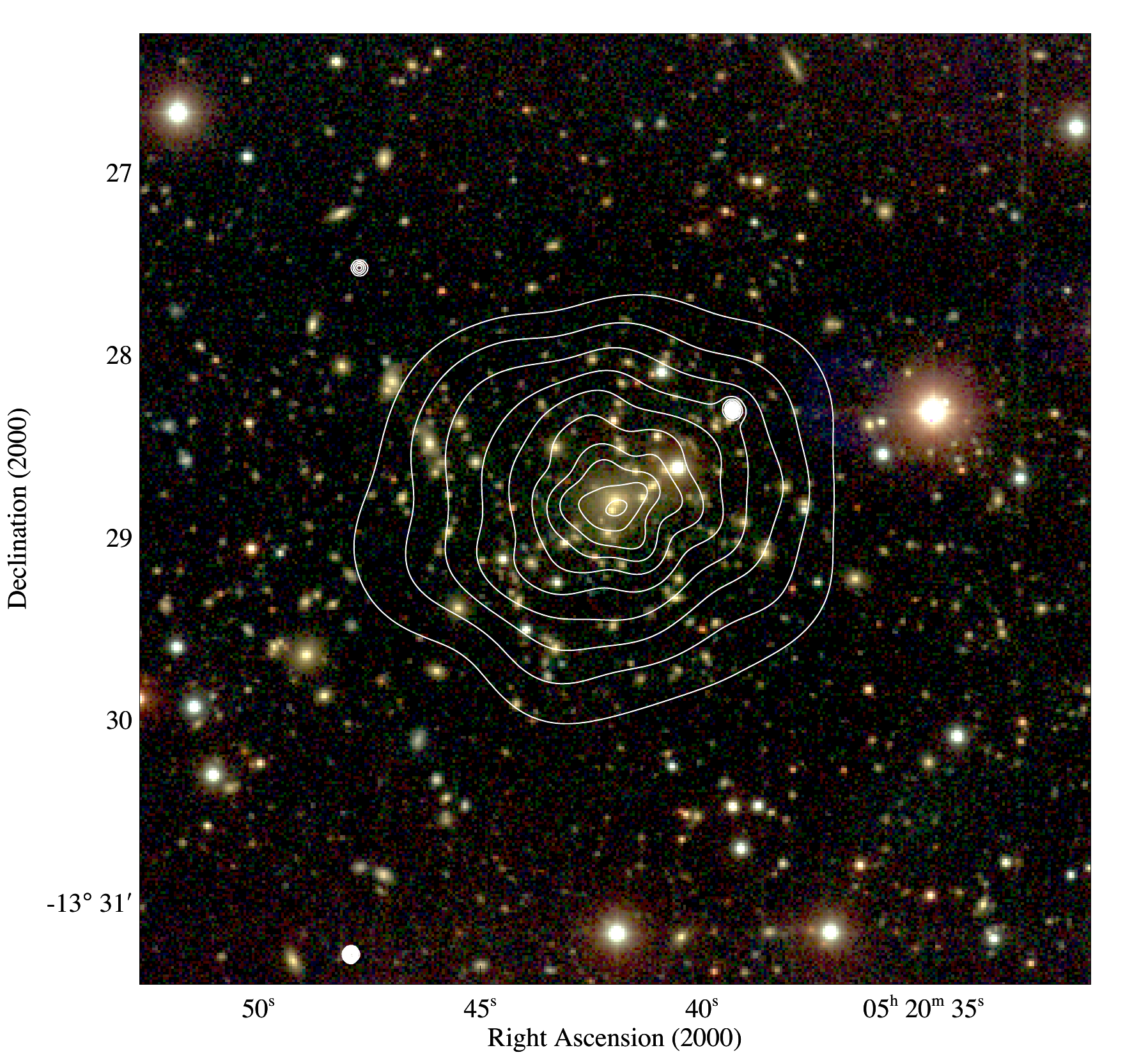}
\includegraphics[type=pdf,ext=.pdf,read=.pdf,width=0.33\textwidth]{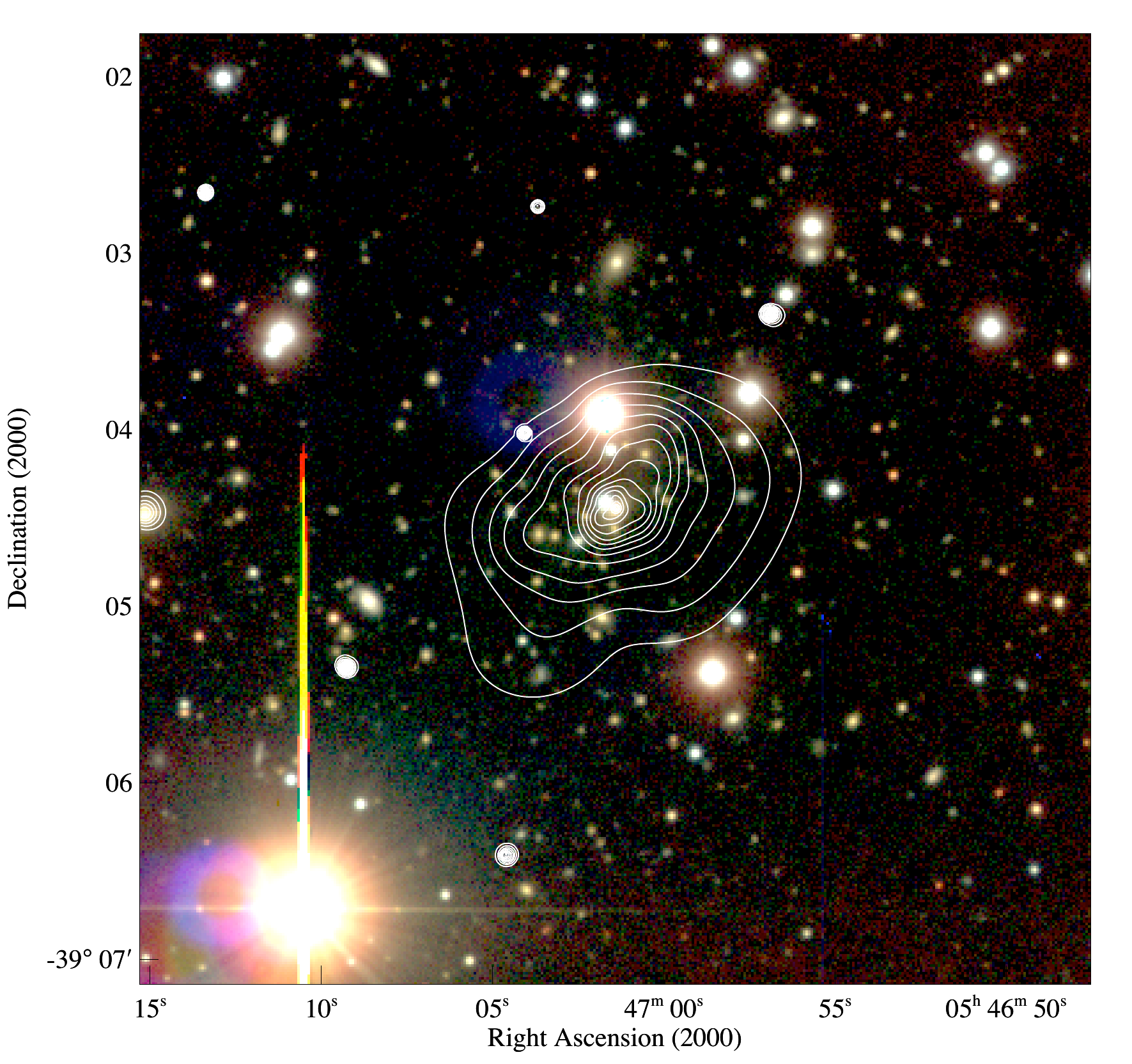}
\includegraphics[type=pdf,ext=.pdf,read=.pdf,width=0.33\textwidth]{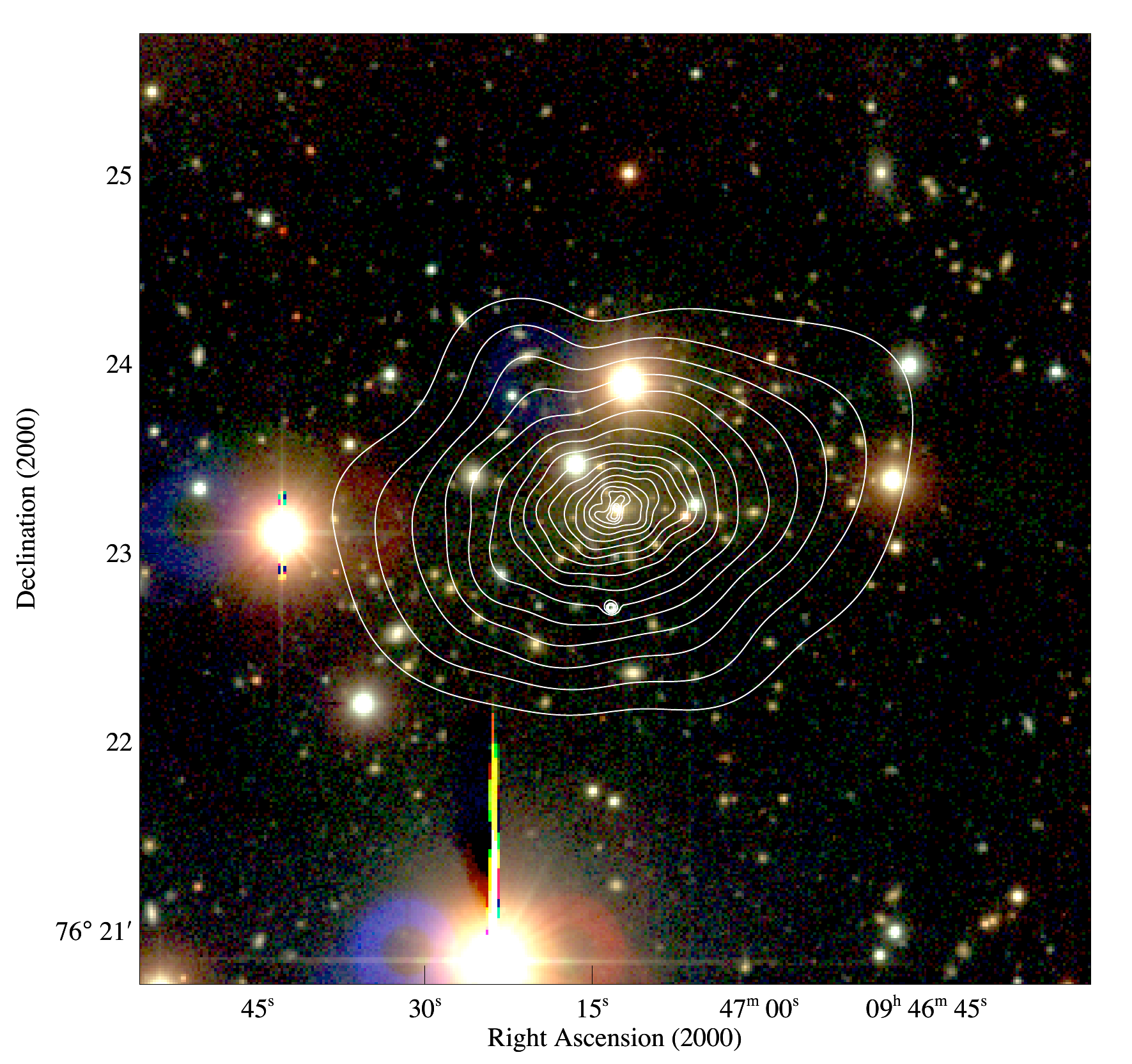}
\includegraphics[type=pdf,ext=.pdf,read=.pdf,width=0.33\textwidth]{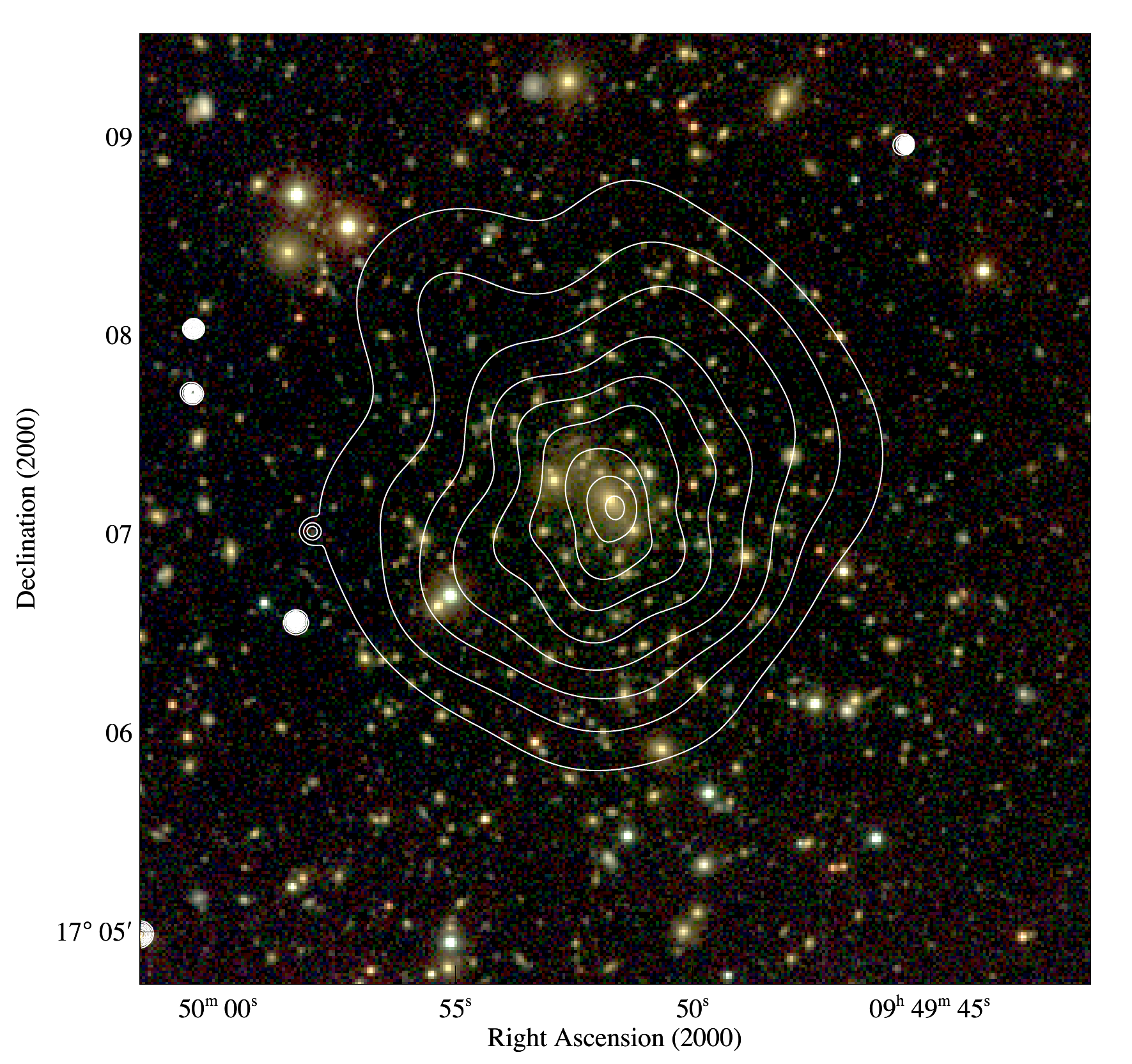}
\includegraphics[type=pdf,ext=.pdf,read=.pdf,width=0.33\textwidth]{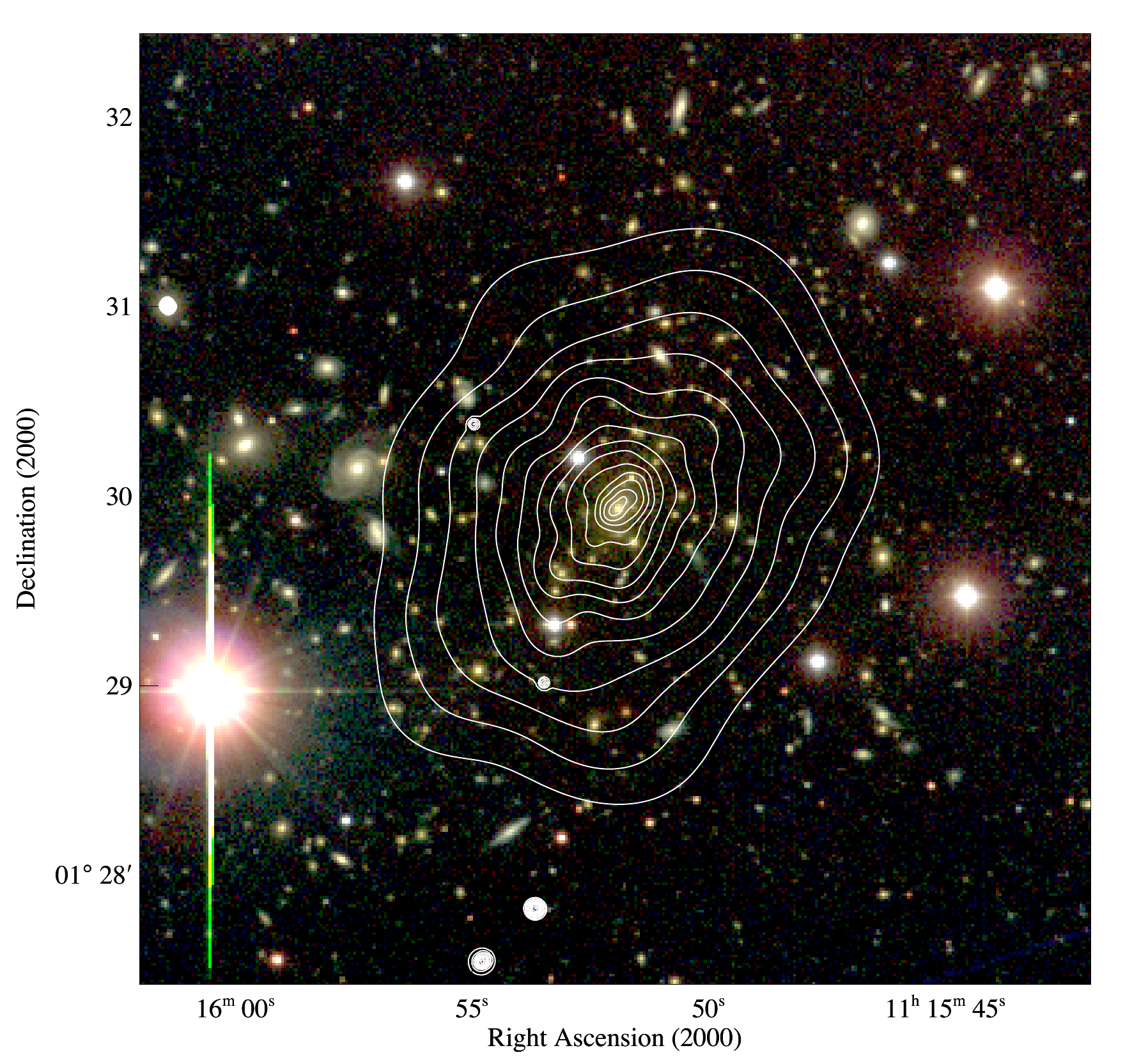}
\includegraphics[type=pdf,ext=.pdf,read=.pdf,width=0.33\textwidth]{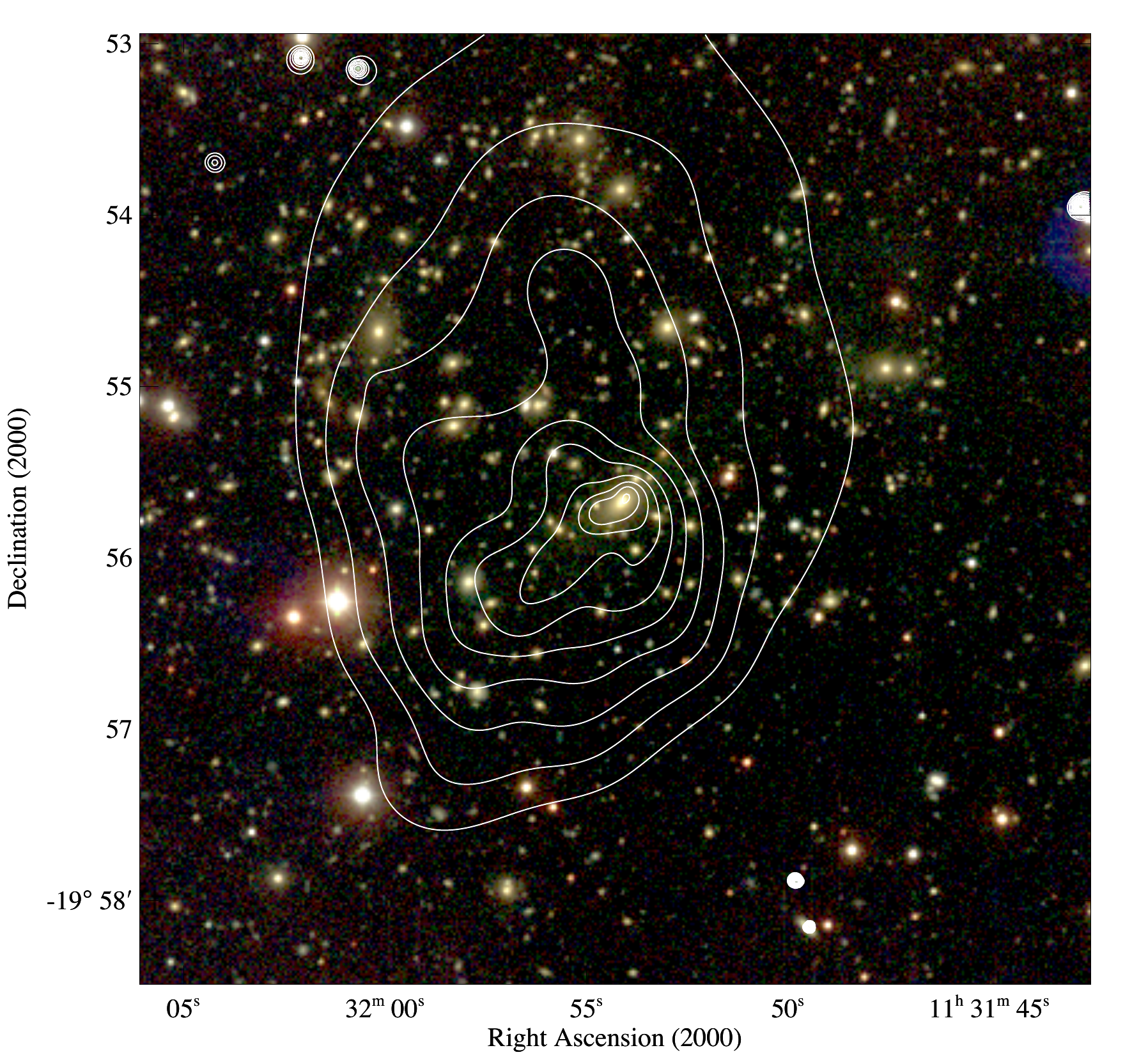}
\includegraphics[type=pdf,ext=.pdf,read=.pdf,width=0.33\textwidth]{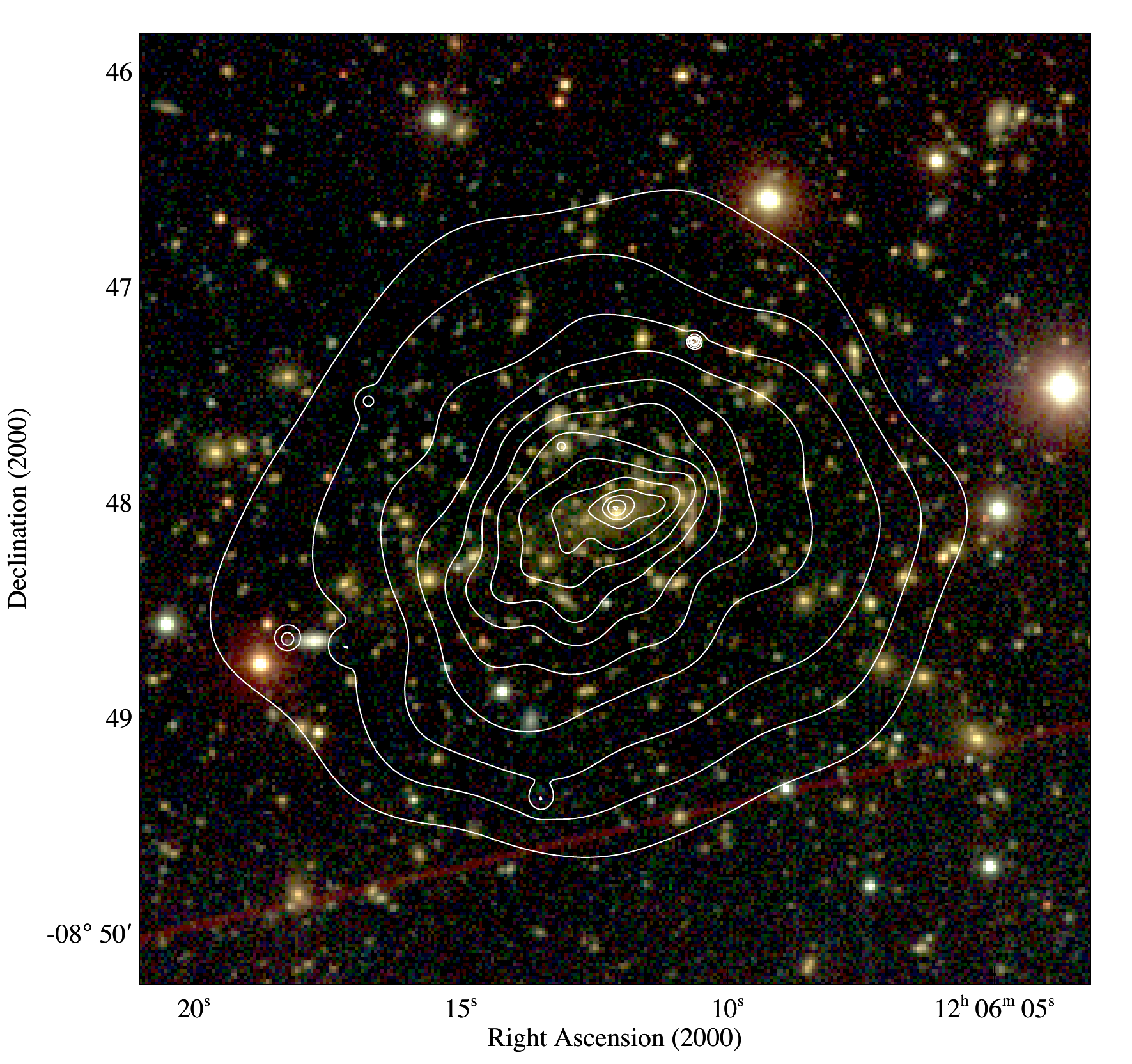}
\includegraphics[type=pdf,ext=.pdf,read=.pdf,width=0.33\textwidth]{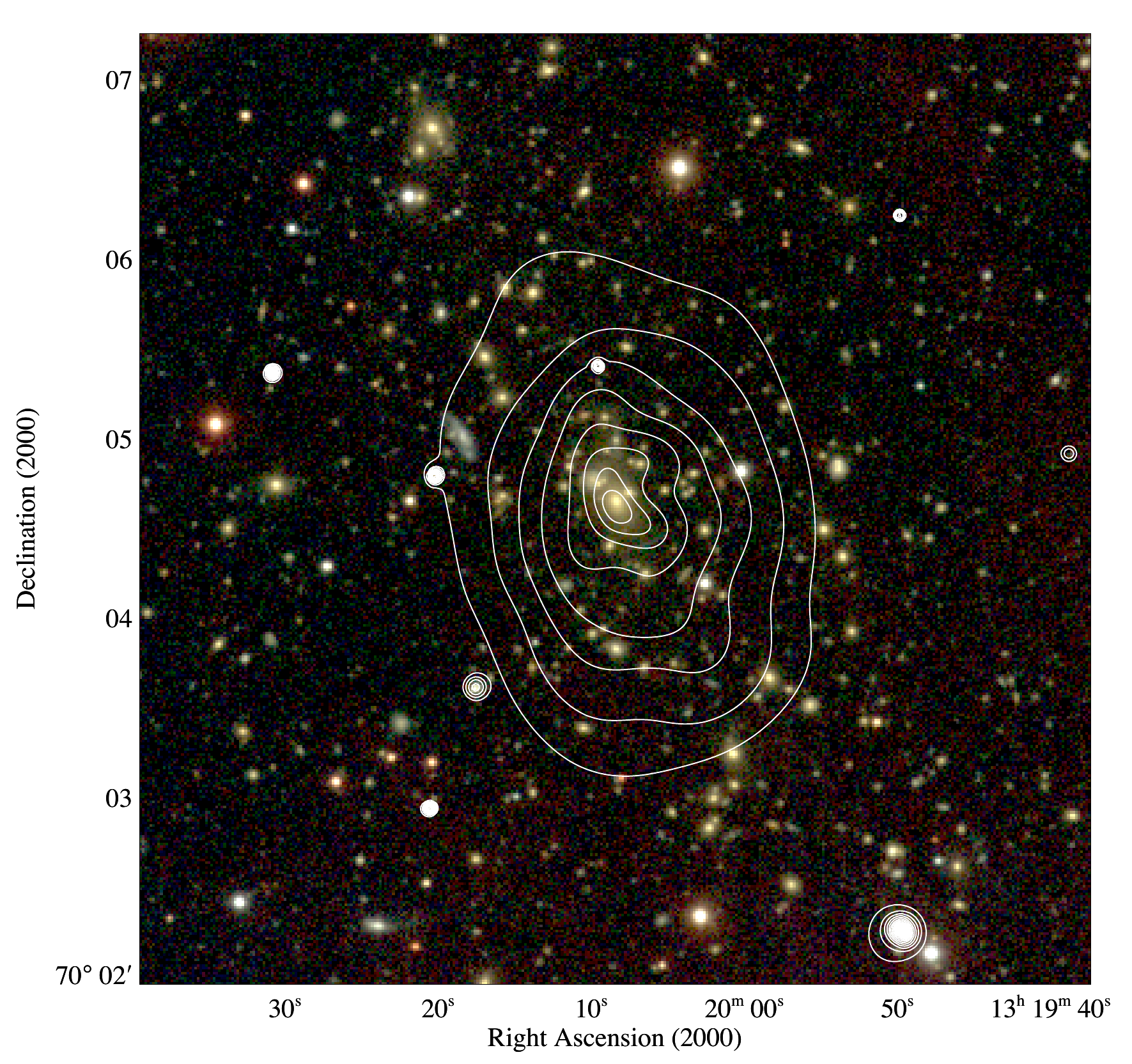}
\includegraphics[type=pdf,ext=.pdf,read=.pdf,width=0.33\textwidth]{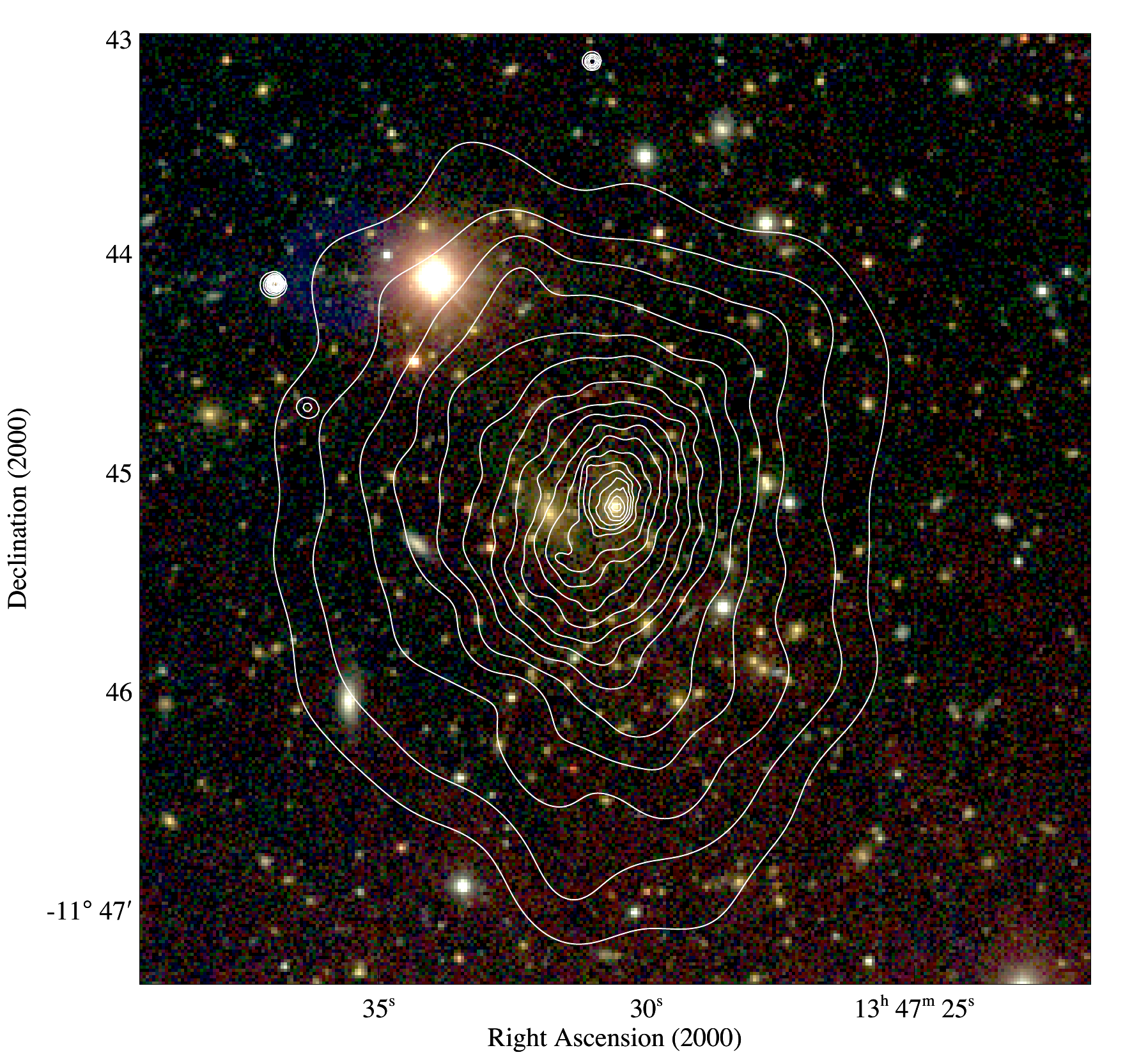}
\includegraphics[type=pdf,ext=.pdf,read=.pdf,width=0.33\textwidth]{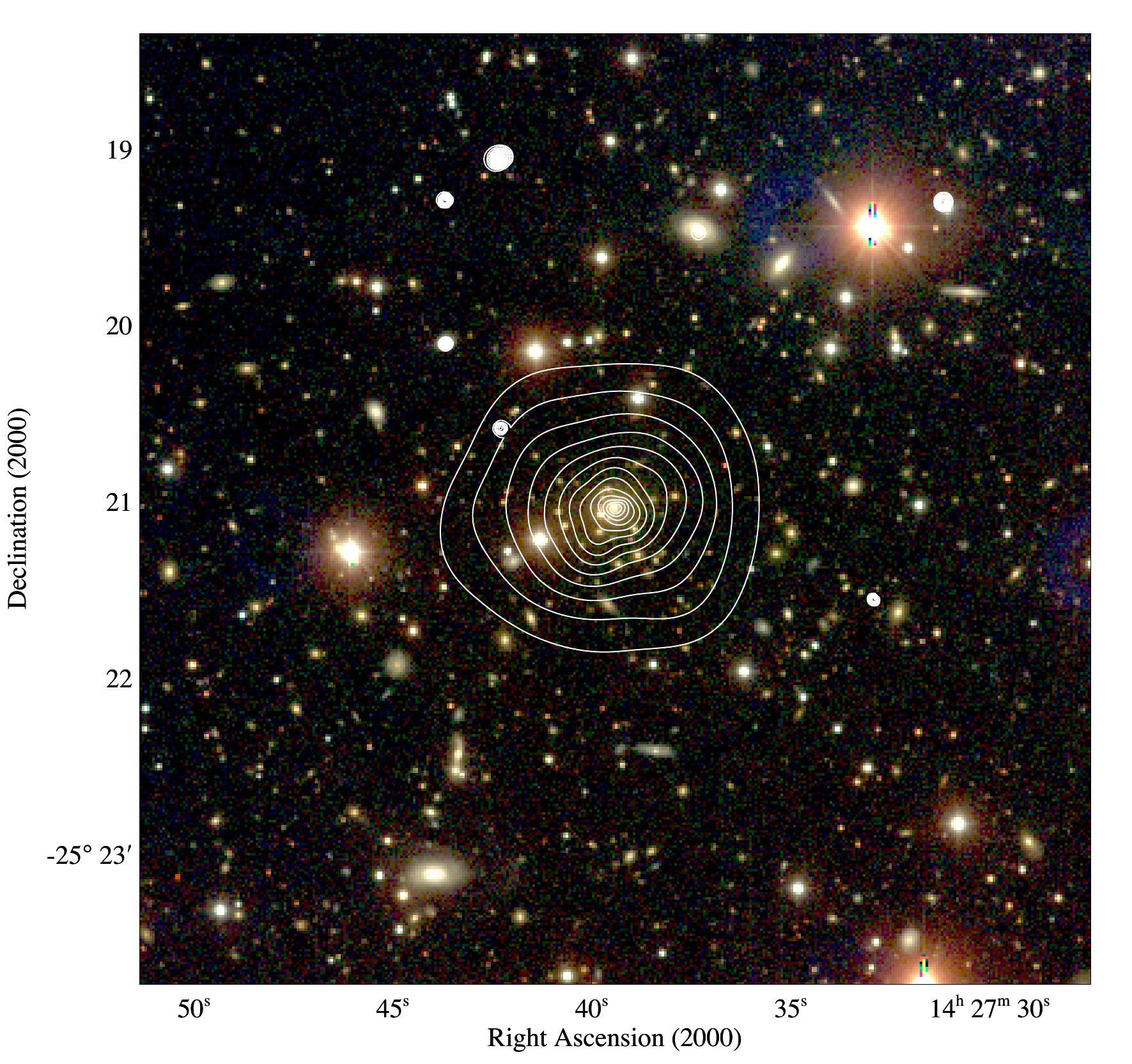}
\includegraphics[type=pdf,ext=.pdf,read=.pdf,width=0.33\textwidth]{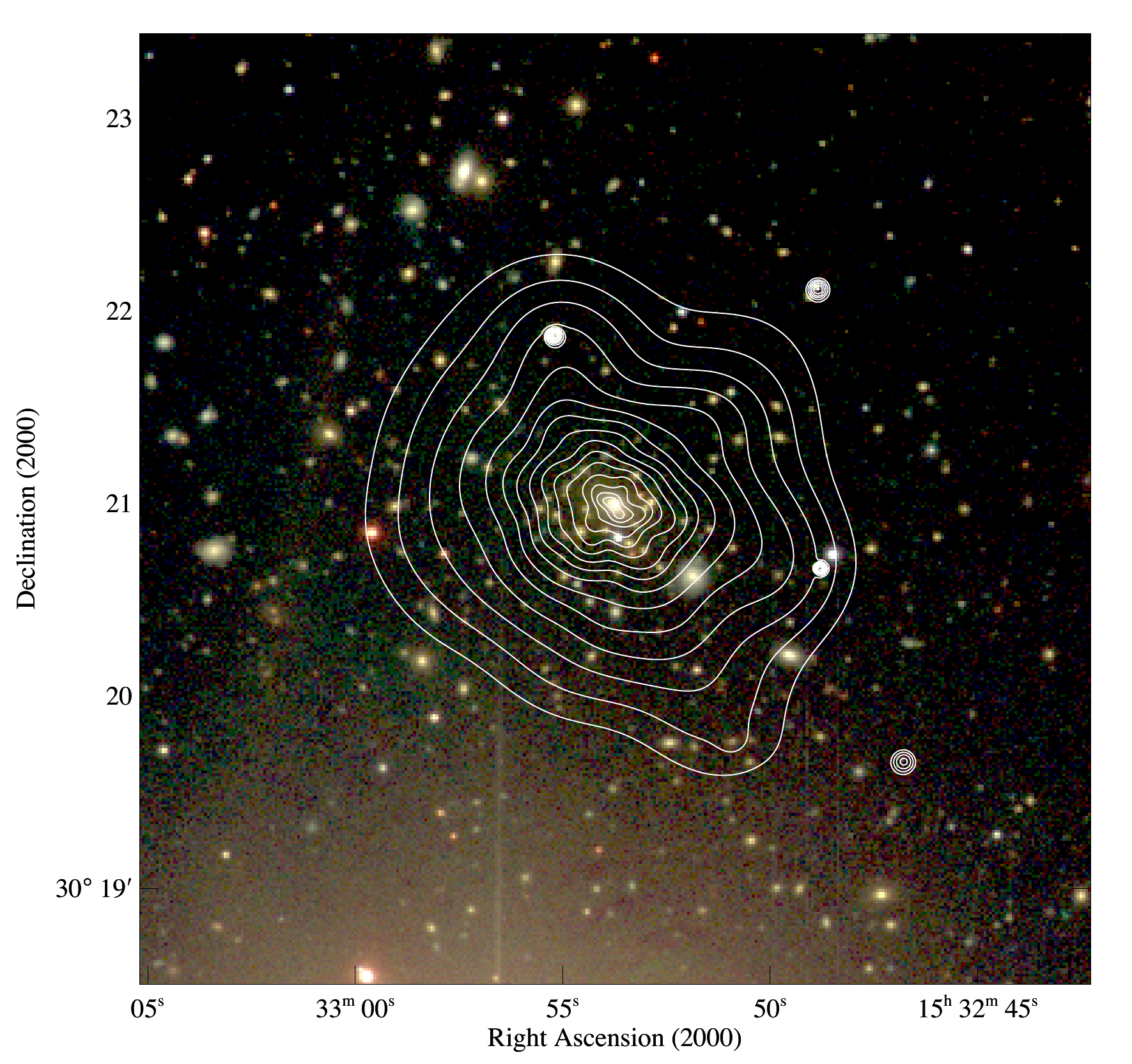}
\includegraphics[type=pdf,ext=.pdf,read=.pdf,width=0.33\textwidth]{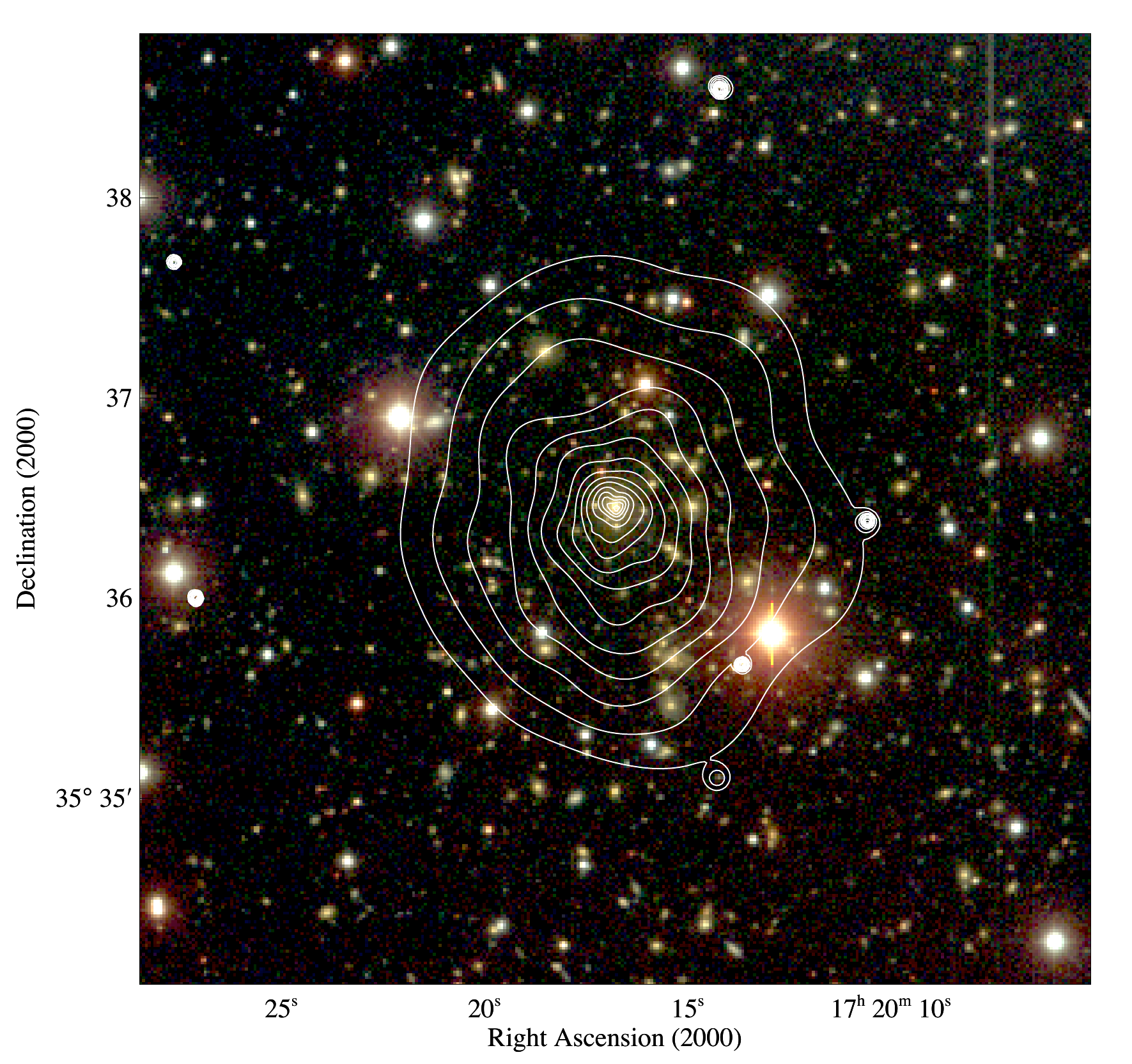}
\contcaption{}
\end{figure*}

\begin{figure*}
\includegraphics[type=pdf,ext=.pdf,read=.pdf,width=0.33\textwidth]{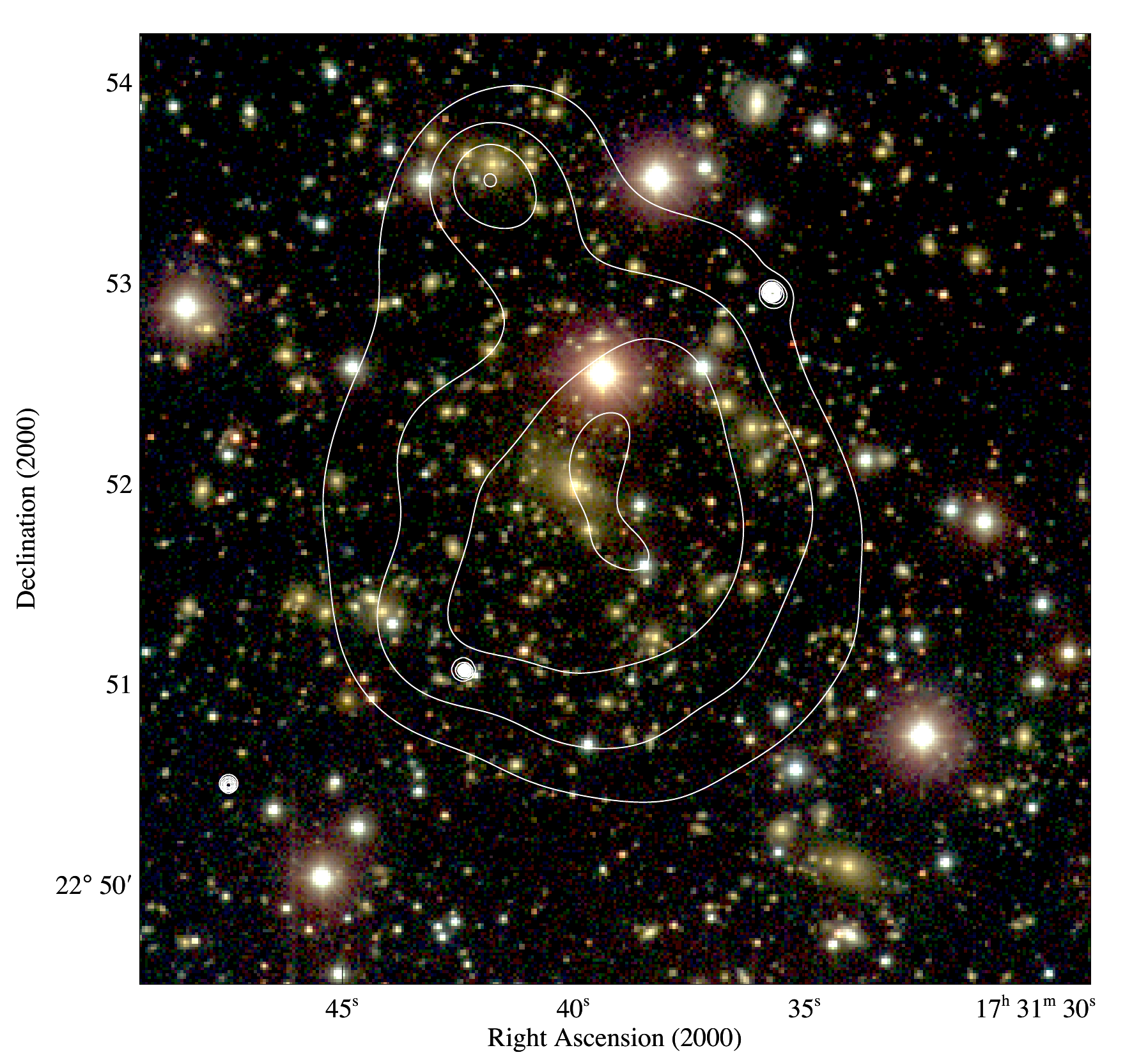}
\includegraphics[type=pdf,ext=.pdf,read=.pdf,width=0.33\textwidth]{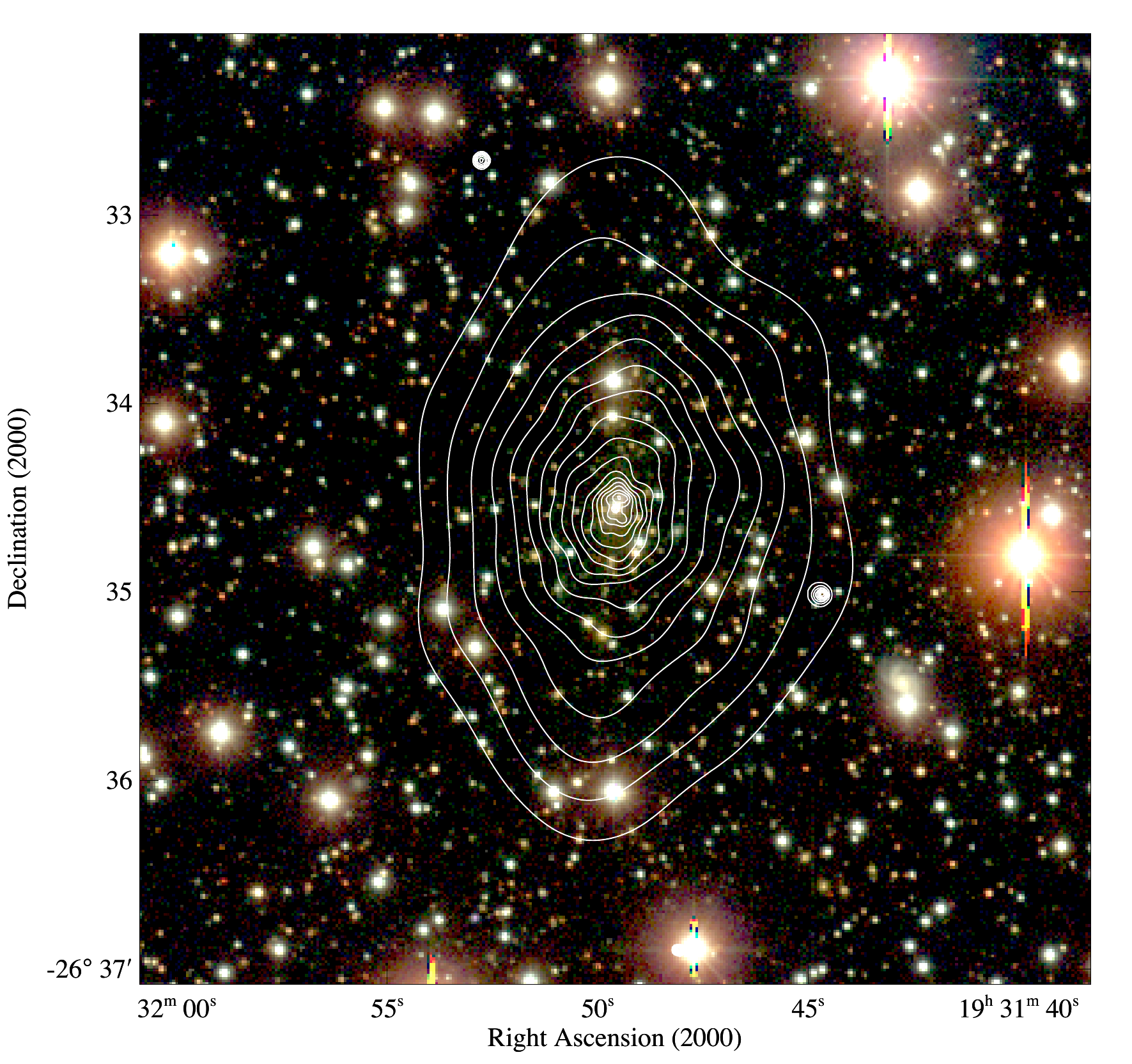}
\includegraphics[type=pdf,ext=.pdf,read=.pdf,width=0.33\textwidth]{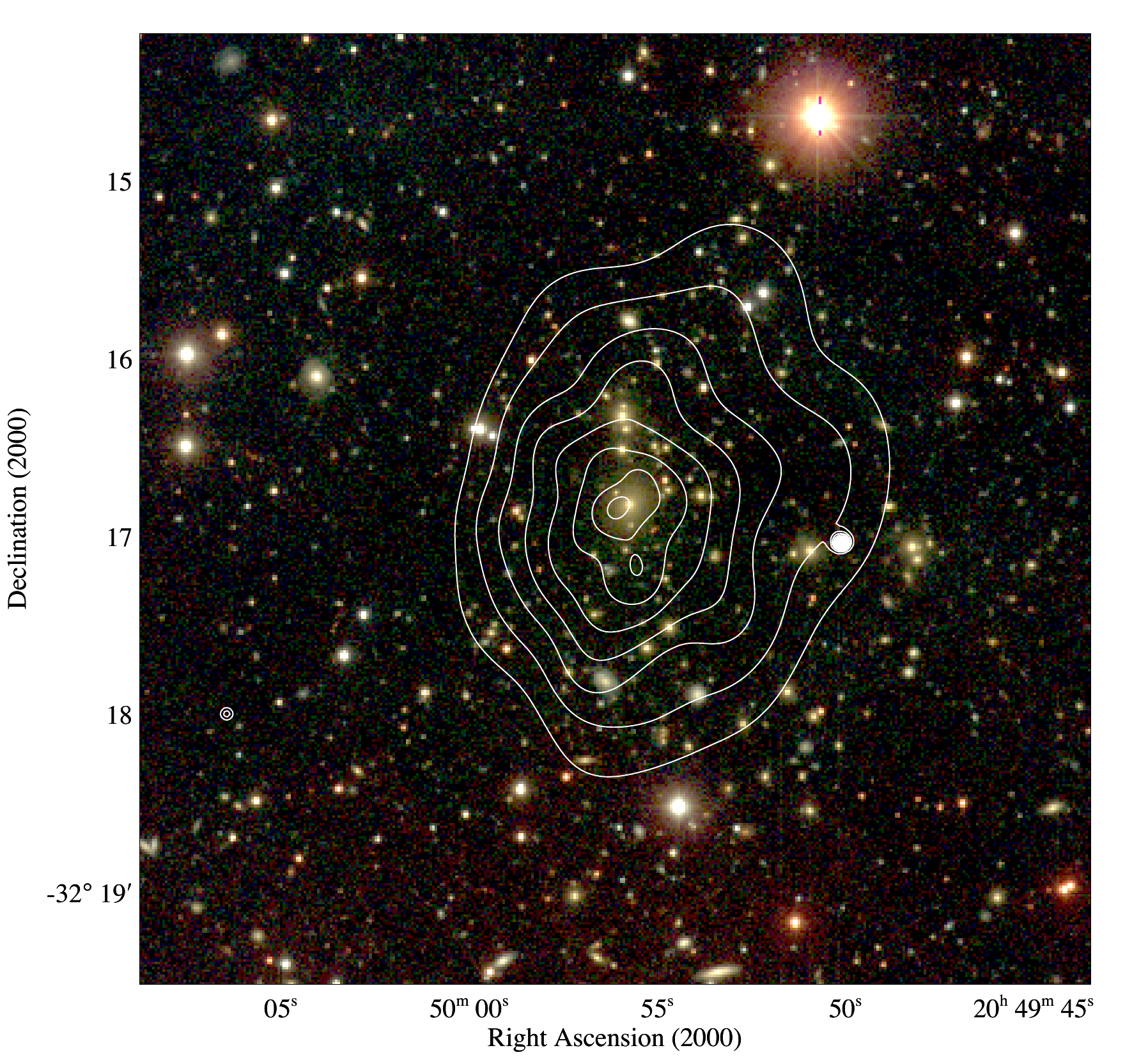}
\includegraphics[type=pdf,ext=.pdf,read=.pdf,width=0.33\textwidth]{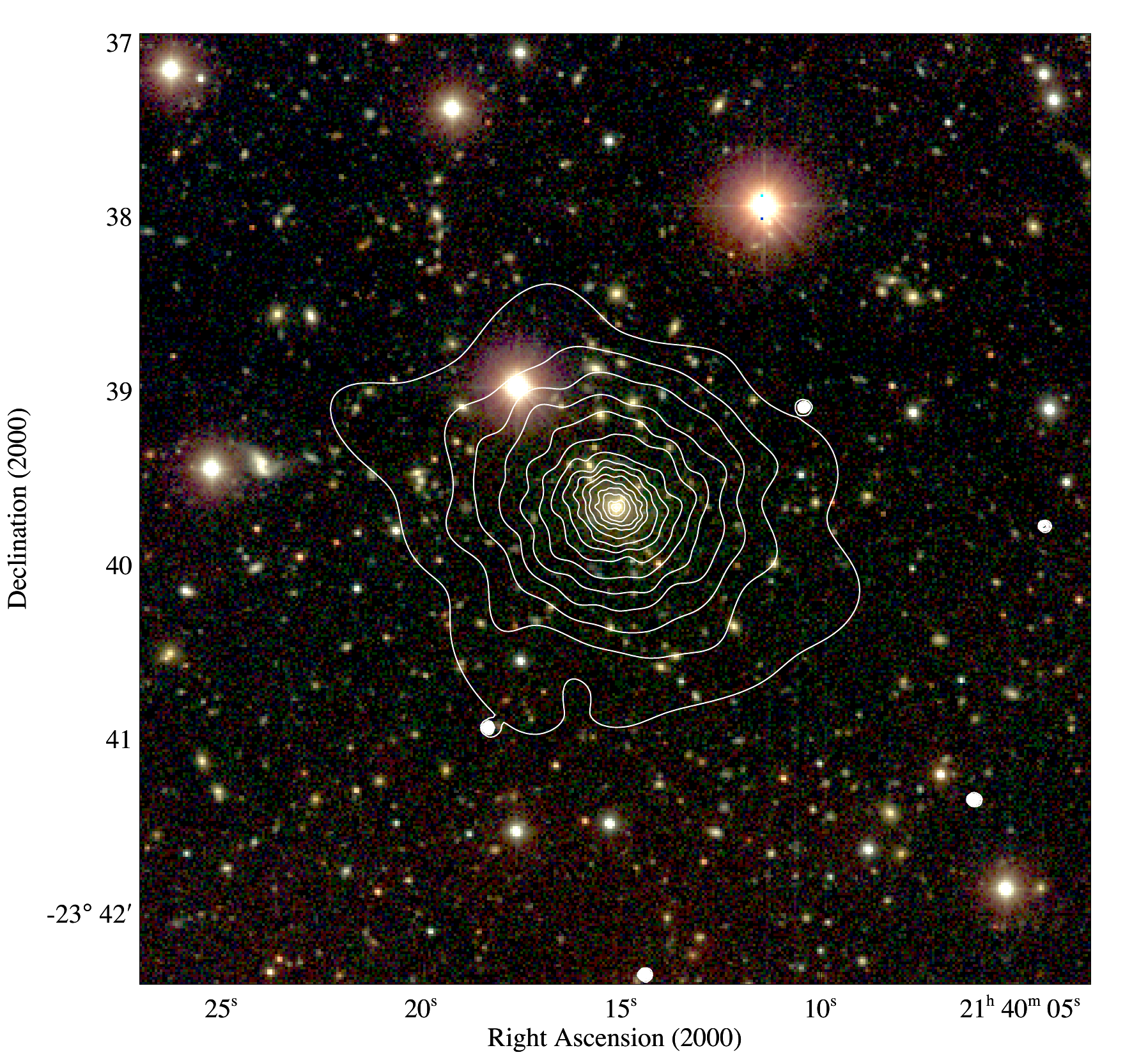}
\includegraphics[type=pdf,ext=.pdf,read=.pdf,width=0.33\textwidth]{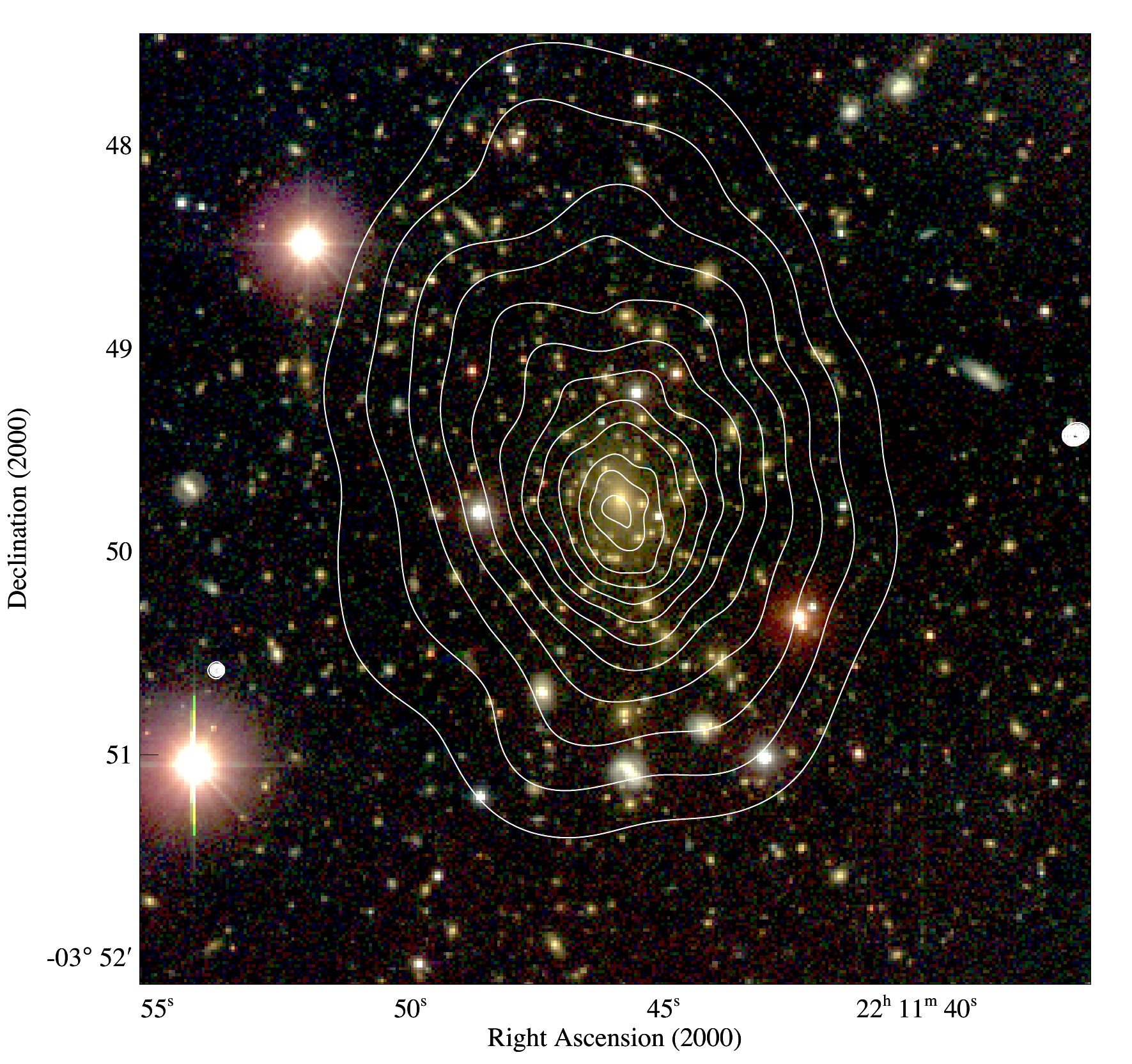}
\includegraphics[type=pdf,ext=.pdf,read=.pdf,width=0.33\textwidth]{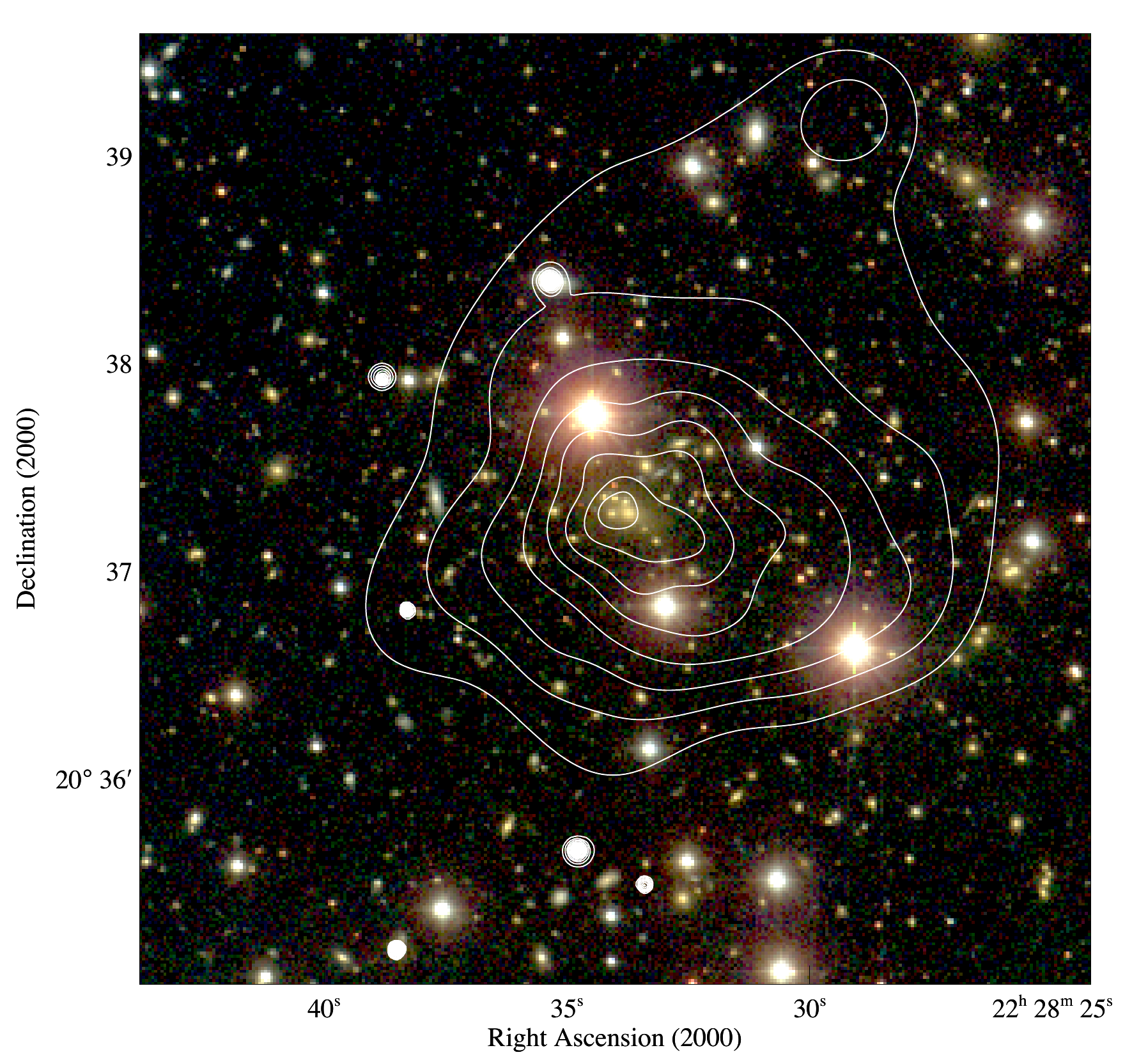}
\includegraphics[type=pdf,ext=.pdf,read=.pdf,width=0.33\textwidth]{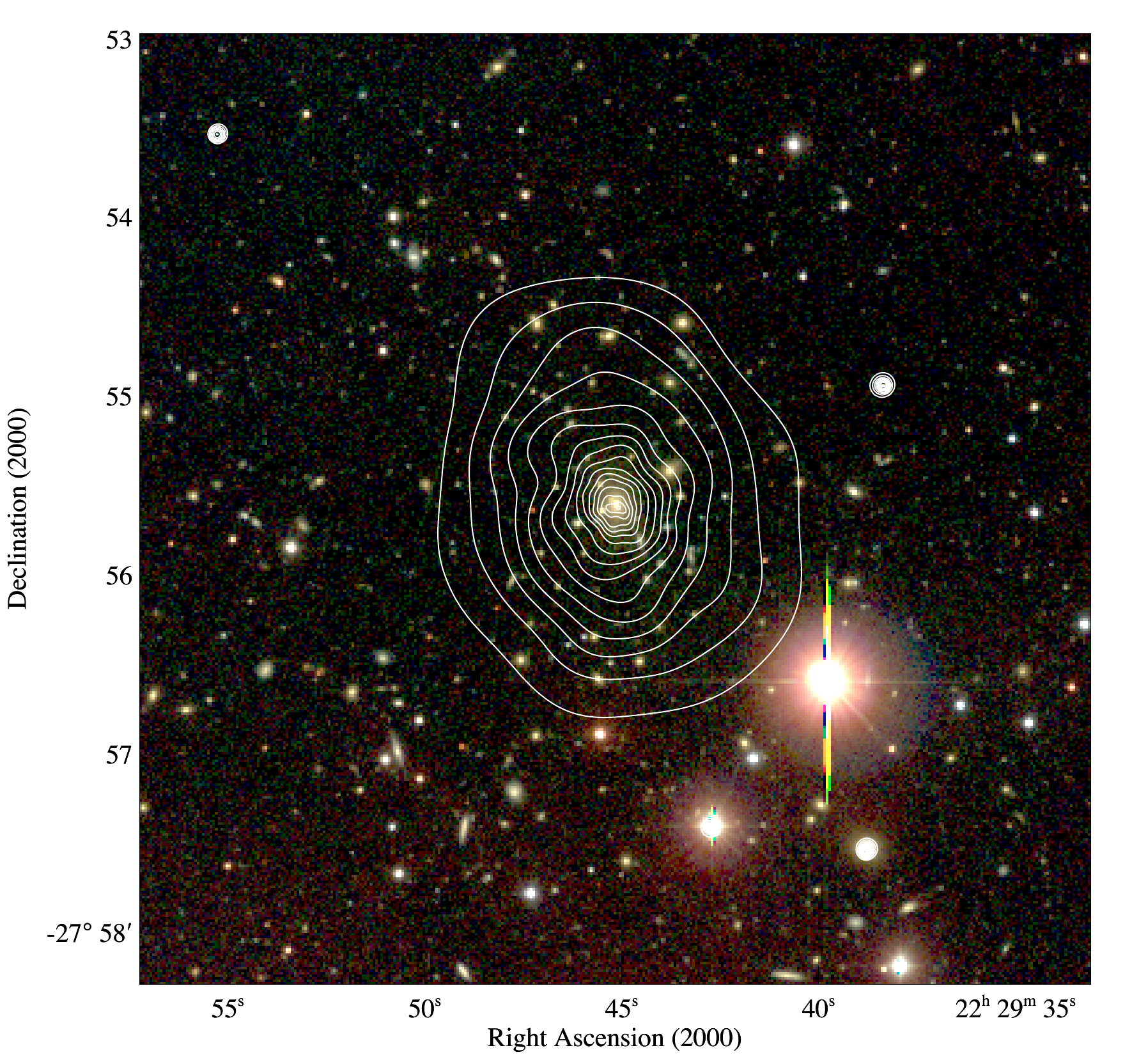}
\includegraphics[type=pdf,ext=.pdf,read=.pdf,width=0.33\textwidth]{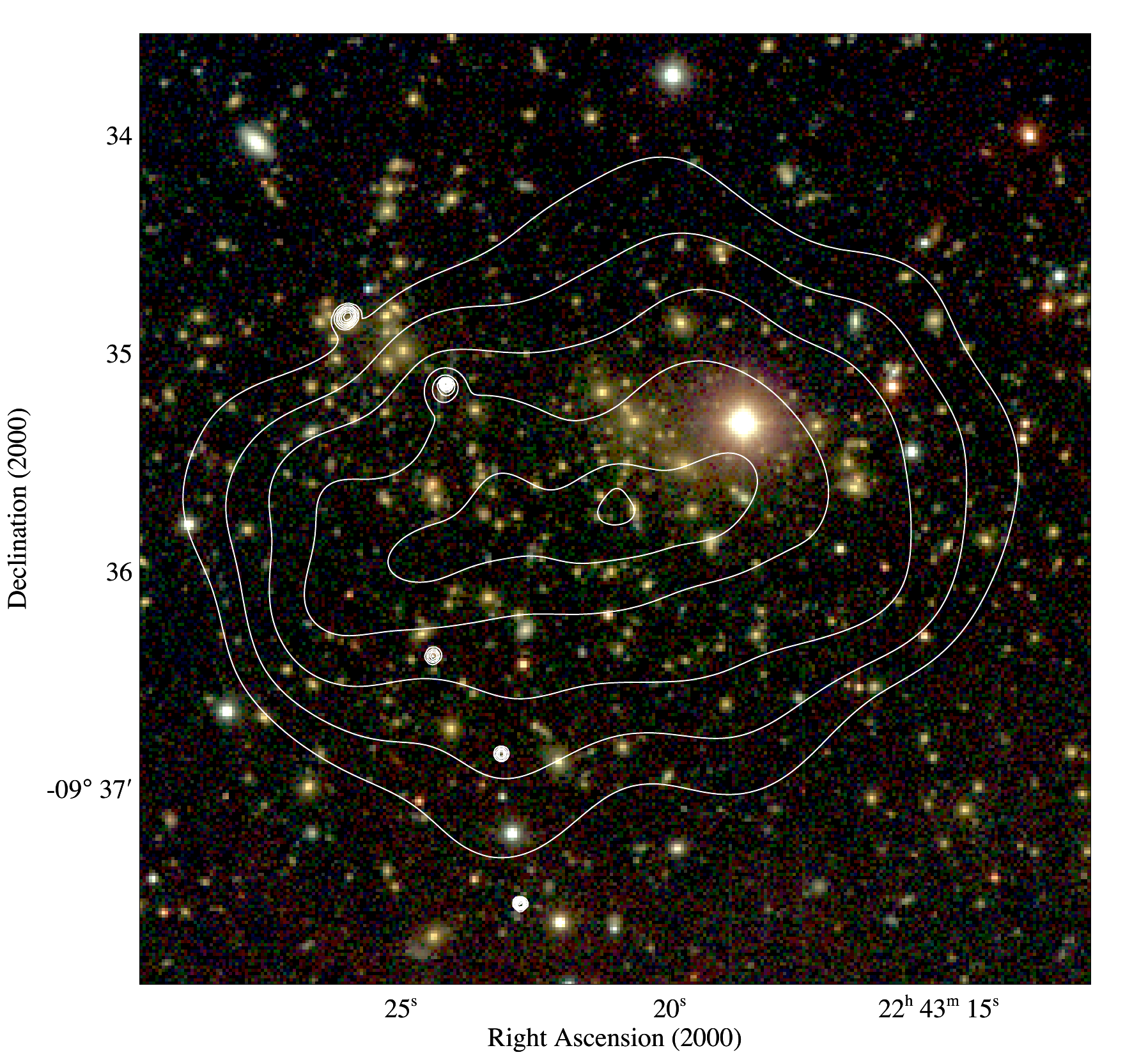}
\includegraphics[type=pdf,ext=.pdf,read=.pdf,width=0.33\textwidth]{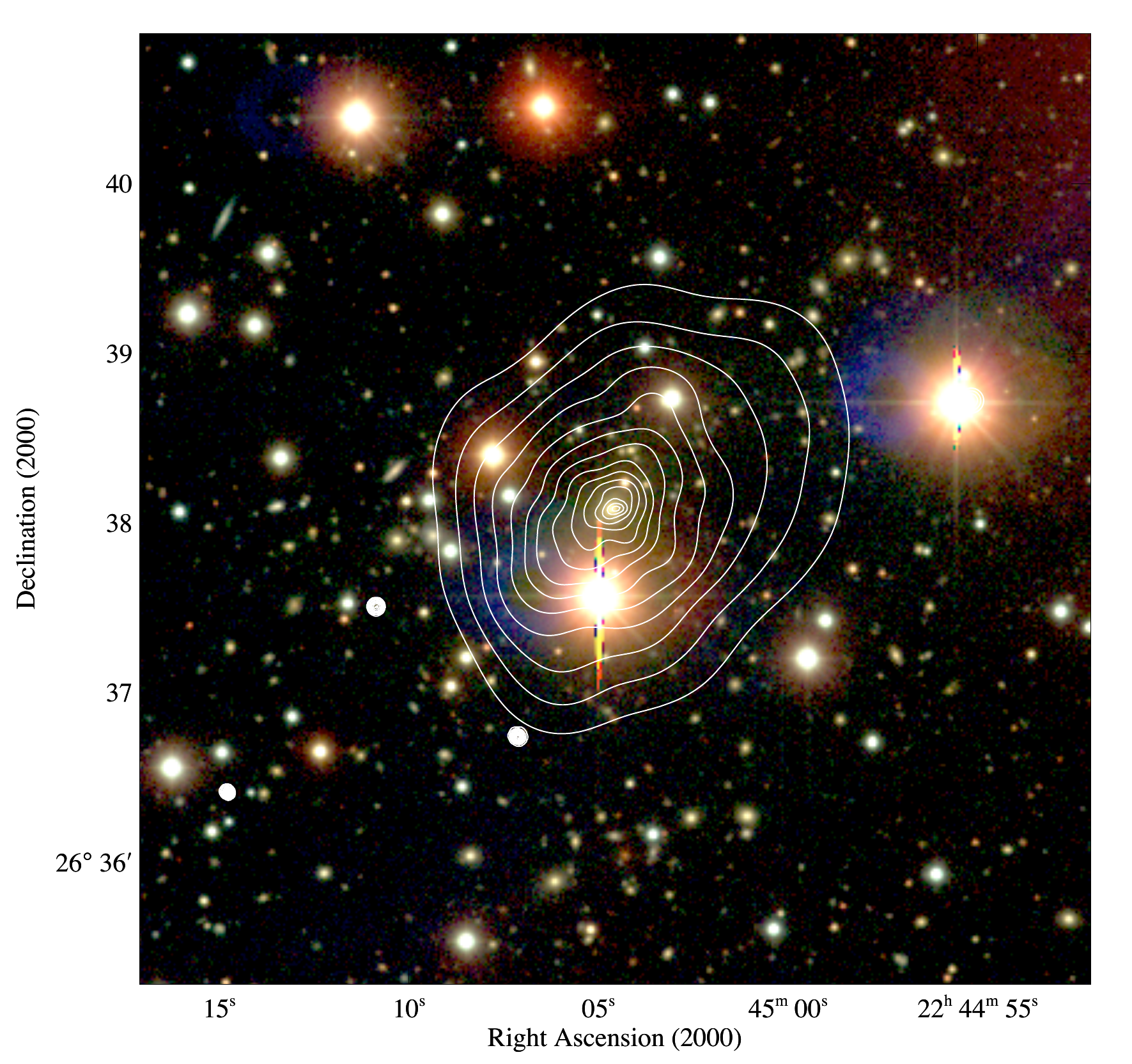}
\includegraphics[type=pdf,ext=.pdf,read=.pdf,width=0.33\textwidth]{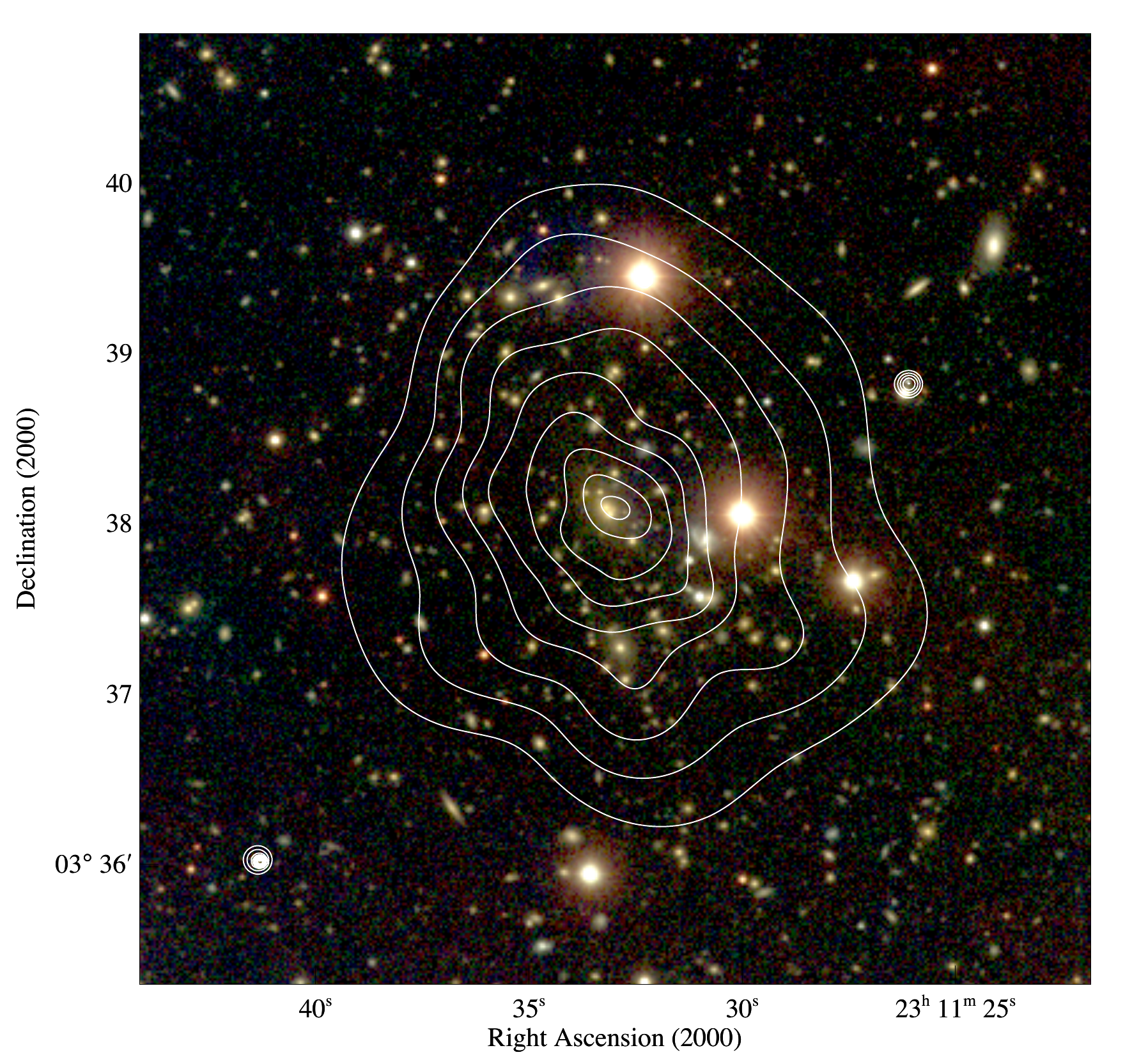} 
\framebox{
\includegraphics[type=pdf,ext=.pdf,read=.pdf,width=0.33\textwidth]{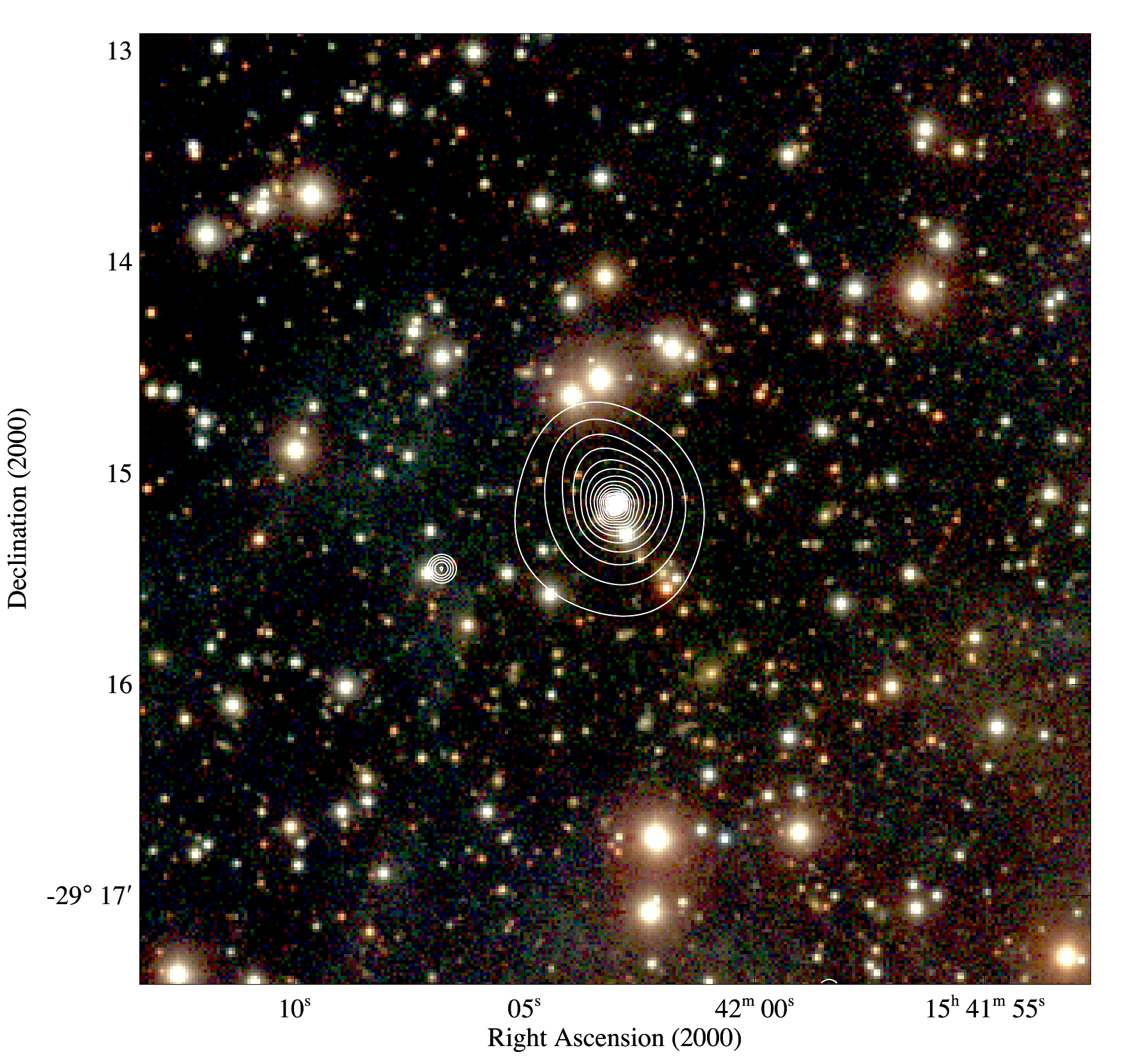}
\includegraphics[type=pdf,ext=.pdf,read=.pdf,width=0.33\textwidth]{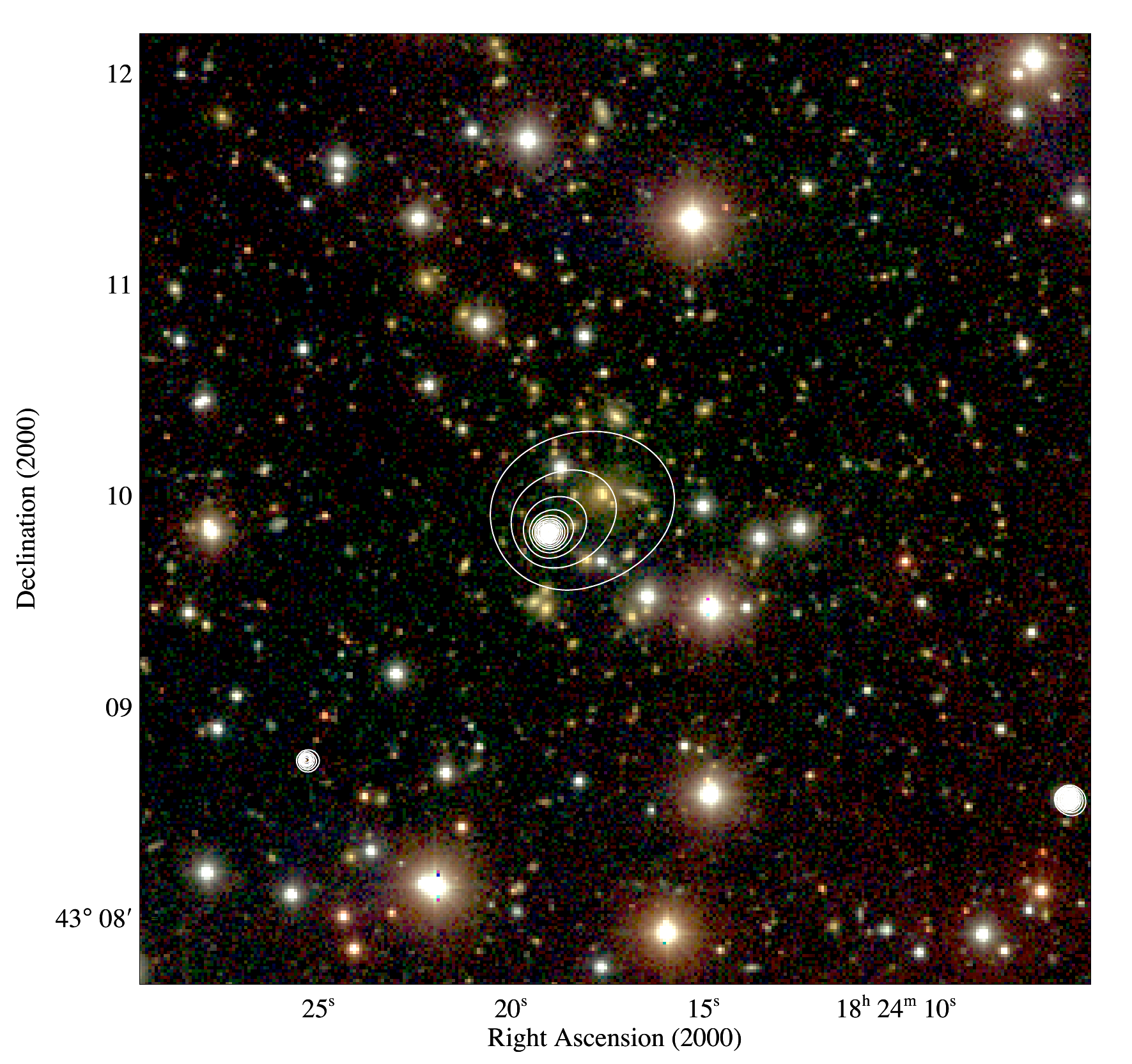}
}
\contcaption{}
\end{figure*}

\subsection{Selection function}

In order to facilitate the use of the presented sample for cosmological applications, we list, in Table~2, the MACS selection function i.e.\ the solid angle covered by our survey as a function of RASS-BSC detect flux (as listed in Table~1). We limit the tabulated range to $f_{\rm det, BSC}\ge 2\times 10^{-12}$ erg
s$^{-1}$ cm$^{-2}$ as the nominal RASS fluxes for yet fainter sources may be affected by 
the systematic effects illustrated in Fig.~4. 

\begin{table}
\caption{MACS selection function. Listed are the RASS detect fluxes $f_{\rm det, BSC}$ in units of $10^{-12}$ erg
s$^{-1}$ cm$^{-2}$ in the 0.1$-$2.4 keV band and the solid angle in square degrees
covered at fluxes exceeding $f_{\rm det, BSC}$.\label{selfun}}
\begin{tabular*}{0.48\textwidth}{@{\extracolsep{-1.5mm}}rr|rr|rr}
\hline 
& & & \\
$f_{\rm det, BSC}$ & solid angle & $f_{\rm det, BSC}$ & solid angle & $f_{\rm det, BSC}$ & solid angle\\ \hline 
& & & \\
      2.00   &    21123 & 5.0      & 22297 & 14.0    &   22533\\
      2.25   &    21432 & 5.5      & 22337 & 15.0    &   22541\\
      2.50   &    21636 & 6.0      & 22369 & 17.5    &   22558\\
      2.75   &    21775 & 6.5      & 22393 & 20.0    &   22569\\
      3.00   &    21886 & 7.0      & 22414 & 22.5    &   22579\\
      3.25   &    21971 & 7.5      & 22430 & 25.0    &   22584\\
      3.50   &    22047 & 8.0      & 22446 & 27.5    &   22589\\
      3.75   &    22116 & 9.0      & 22469 & 30.0    &   22594\\
      4.00   &    22167 & 10.0    & 22488 & 35.0    &   22601\\
      4.25   &    22213 & 11.0    & 22502 & 40.0    &   22607\\
      4.50   &    22246 & 12.0    & 22514 & 45.0    &   22614\\
      4.75   &    22277 & 13.0    & 22525 & 50.0    &   22617\\ \hline
\end{tabular*}
\end{table}

\section{Summary}

We present the second statistically complete MACS subsample, comprising the 34
clusters with X-ray detect fluxes in excess of $2\times 10^{-12}$ erg s$^{-1}$
cm$^{-2}$ (0.1--2.4 keV) in the RASS Bright Source Catalogue. All clusters
feature redshifts of $z\ge 0.3$, and 22 of the 34 are new discoveries. Chandra
observations of the entire sample confirmed the cluster origin of the emission
and allowed the elimination of three additional candidates whose X-ray emission
was found to be dominated by point sources.

A comparison of the appearance of MACS clusters in the RASS and in Chandra
observations confirms that all but the most disturbed clusters at $z>0.3$ appear
point-like at the angular resolution of the RASS. We find the original RASS
count rates as listed in the BSC to be less accurate than manual measurements
within the BSC aperture and using a local annulus for background
subtraction. RASS fluxes based on recomputed count rates are in good agreement
with the respective Chandra values, except for four clusters for which the RASS
count rate is significantly contaminated by point sources within the BSC detect
cell, and one system whose very extended emission is not fully captured by the
Chandra measurement. For the remainder of the sample, X-ray point sources 
contribute, on average, only about 3\% to the flux within $r_{\rm 500}$. All clusters of this second MACS subsample feature X-ray
luminosities (within $r_{\rm 500}$) in excess of $4\times 10^{44}$ erg s$^{-1}$
(0.1--2.4 keV) after correction for X-ray point sources, and are thus
considerably more X-ray luminous than the Coma cluster. The sample's median
X-ray luminosity of $1.3\times 10^{45}$ erg s$^{-1}$ confirms the efficiency of
our survey technique to identify massive clusters well beyond the redshift
limits of previous RASS-based cluster surveys. A first assessment of the optical
and X-ray morphology of the clusters in this sample finds both fully virialized
and heavily disturbed systems to be well represented, arguing against a strong
bias in favour of either cool-core clusters or extreme mergers. A more quantitative
analysis and discussion of the morphology and relaxation state of MACS clusters
will be presented in a forthcoming paper.

When combined with the most X-ray luminous clusters in the local Universe
($z<0.3$) from the eBCS and REFLEX surveys (Ebeling et al.\ 1998, 2000;
B\"ohringer et al.\ 2004) and the complete set of the 12 most distant MACS
clusters ($z>0.5$) released earlier (Ebeling et al.\ 2007), the sample presented
here allows cosmological and astrophysical studies of the properties and
evolution of the most massive galaxy clusters over a contiguous redshift range
from $z\sim 0$ to $z=0.7$ (Allen et al.\ 2003, 2004, 2008; Ebeling et al.\ 2009;
Mantz et al.\ 2008, 2010a,b, Rapetti et al.\ 2009). For convenience, we have supplied 
a tabulated version of the appropriate selection function.

\section*{Acknowledgments}

We thank many incarnations of the Chandra peer-review panel and of the
University of Hawaii's telescope time allocation committee for their support,
trust, and patience.  HE gratefully acknowledges financial support from NASA
LTSA grant NAG 5-8253 and SAO grants GO2-3168X, GO5-6133X, and GO0-11140X. ACE
thanks the Royal Society for generous support during the identification phase of
the MACS project. Parts of this work received support from the U.S. Department
of Energy under contract number DE-AC02-76SF00515, as well as from SAO grants
DD5-6031X, GO7-8125X and GO8-9118X. AM was supported by a Stanford Graduate
Fellowship and an appointment to the NASA Postdoctoral Program, administered by
Oak Ridge Associated Universities through a contract with NASA.

\end{document}